\documentclass[journal,10pt,twocolumn,twoside]{IEEEtran} 
\IEEEoverridecommandlockouts
\usepackage{amsmath}
\usepackage{amsfonts}
\usepackage{cases}
\usepackage{setspace}
\usepackage{fancybox}
\usepackage{subfigure}
\usepackage{epsfig}
\usepackage{graphicx}
\usepackage{epstopdf}
\usepackage{float}
\usepackage{multirow}
\usepackage{color}%
\usepackage{amsmath}
\usepackage{multirow}

\usepackage{indentfirst}
\usepackage{dsfont}
\usepackage{amsfonts}
\usepackage{times,amsmath,color,amssymb,epsfig,cite,subfigure,algorithm,algorithmic}
\usepackage{array}

\usepackage{booktabs}

\newcommand{\qed}{\nobreak \ifvmode \relax \else
      \ifdim\lastskip<1.5em \hskip-\lastskip
      \hskip1.5em plus0em minus0.5em \fi \nobreak
      \vrule height0.40em width0.6em depth0.25em\fi}

\newtheorem{lemma}{Lemma}
\newtheorem{theorem}{Theorem}

\begin{document}
%
\title{Designing Binary Sequence Set with Optimized Correlation Properties via ADMM Approach}

\author{

Jiangtao~Wang,
 }

%

\markboth{}%
{}

\maketitle

\begin{abstract}
In this paper, we design low correlation binary sequences favorable in wireless communication and radar applications.
First, we formulate the designing problem as a nonconvex combination optimization problem with flexible correlation interval;
second, by relaxing  constraints and introducing auxiliary variables, the original minimization problem is equivalent to a consensus continuous optimization problem;
third, to achieve its good approximate solution efficiently, we propose the distributed executable algorithms based on alternating direction method of multipliers (ADMM);
fourth, we prove that the proposed ADMM algorithms can converge to some stationary point of the approximate problem.
Moreover, the computational complexity analysis is considered.
Simulation results demonstrate that the proposed ADMM approaches outperform state-of-the-art ones in either computational cost or selection of correlation interval of the designed binary sequences.
\end{abstract}

\begin{IEEEkeywords}

Binary sequence, auto/cross-correlation, Box ADMM, convergence/complexity analysis.

\end{IEEEkeywords}

\IEEEpeerreviewmaketitle

\section{Introduction}

\IEEEPARstart{C}ONSTANT modulus sequences with good correlation play a fundamental role in modern communication systems \cite{Chu_72}--\hspace{-0.001cm}\cite{Velazquez_17}. These sequences are widely used in a variety of applications, such as power control \cite{Liu_14}, channel estimation \cite{Milewski_83}, \cite{Choi_14}, synchronization \cite{Lambrette_97}, signal detection and separation \cite{Golay_61}, and mitigation of interference \cite{Tang_06}, etc.
Among these sequences, binary sequences, such as $m$-sequences \cite{Golomb_82}, Kasami code \cite{Lahtonen_95}, Gold code \cite{Gold_68}, Barker code \cite{Barker_53}, Bent code \cite{Kumar_83}, etc., have been widely studied due to their simplicity of implementation and high-energy efficiency.

Generally speaking, perfect correlation sequences are desired in communication systems when their autocorrelation values are zero for all shifts different from zero and corss-correllation values are zero for every shift. It is known that these perfect unitary sequences do not exist and this is the main challenge to construct zero-correlation sequences limited by bounds \cite{Velazquez_17}.
For Baker codes, Storer and Turyn proved that there are no sequences  for odd $N \geq 13$ \cite{Turyn_61} and except $N=4$, no other perfect period binary sequences exist with $N<548964900$ \cite{Leung_05}.
Therefore,  zero correlation zone (ZCZ) sequence, whose correlation values within a zone are zero, are proposed in \cite{Fan_99} and  and so do low correlation zone (LCZ) sequences in \cite{Chung_2010}.
In fact, the construction of low correlation binary sequences is a well-known computational problem.

At the early stage, exhaustive search method is the main tool in  the construction of short sequence with ideal correlation properties.
The authors in \cite{Klapper_93} used the trace function to construct cascaded Gordon-Mills-Welch (GMW) sequence with low autocorrelation and three valued cross-correlation.
The authors in \cite{No_98} presented five new classes of binary sequences of period $2^N-1$ with ideal autocorrelation by extensive computer search.
Based on the framework of exhaustive search algorithm, authors in \cite{Mertens_96} customized the fast method to construct binary sequences with low autocorrelations.
In \cite{Deng_04}, the author adopted a hybrid approach combining simulated annealing approach with a traditional iterative code selection algorithm to design orthogonal polyphase sequence sets.
The authors in \cite{Wang_08} proposed an iterated variable depth search algorithm to search binary sequences with integrated sidelobe level (ISL) and low peak sidelobe level (PSL).
Owing to the exponential size $\mathcal{O}(2^N )$ of the configuration space, above exhaustive search methods are limited to design short sequences.

Pseudo-Noise (PN) sequences generated from Feedback Shift Registers (FSR) can be designed as long sequences.
A kind of these sequences, $m$-sequences, are easily generated using linear FSR with length $2^N - 1$.
Many sequences, such as Gold code, Kasami code, GMW sequence are derived from $m$-sequences.
But the disadvantage of these sequences is that they are relatively small in number \cite{Warty_13}.
Interleaved technique is another method used to analyse and design  sequences with good correlation \cite{Gong_95}.
Its key idea is construct long sequences from short ones \cite{Li_11},\cite{Tang_10}.
More literature on this area can be found in \cite{Yu_08}-\hspace{-0.01cm}\cite{Su_18} and the references therein.
This kind of computational design method is simple to implement, but it can only be designed for sequences with length of $2^N-1$ or $2^N$, which leads to its lack of flexibility.

The limitations motivated the researchers to perform analytical construction method to design sequences.
In recent years, a large number of literatures related to constant modulus sequences design have emerged (see \cite{Song_15}--\hspace{-0.001cm}\cite{Wang_19} and references therein).
 {As a special case of that, the study of discrete sequence also attracts many researchers.}
The authors in \cite{Bose_18} bridged the gap between the exhaustive search method and the analytical constructions method, and proposed a construction method which can be accomplished in polynomial time.
The authors in \cite{Mo_20} formulated the sequence sets design as a quadratically constrained quadratic program problem and proposed an algorithm based on semidefinite program (SDP) relaxation and randomized projection technique to tackle it.
An efficient coordinate-descent framework method was proposed to design sequences with good ISL/PSL for MIMO radars and communication systems in \cite{Kerahroodi_17},\cite{Kerahroodi_19}.
Due to element-wise optimization, this method has the drawback of low efficiency.
In \cite{Cui_17},\cite{Yu_20}, the authors formulated the sequence design with similarity constraint as quadratic optimization problem, the Serial Iterative Algorithm (SIA) \cite{Cui_17} and Alternating Direction Penalty Method (ADPM) algorithm \cite{Yu_20} are proposed successively.
The authors in \cite{Zhang_20} adopt the effective alternating direction method of multipliers (ADMM) to design discrete phase sequence with desired minimized sidelobes.
Majorization-Minimization (MM) technique is also considered to tackle the quartic objective function. However, this method has the disadvantage of relatively high computational complexity.


 In this paper, we focus on designing binary sequence sets with good correlation properties with low complexity algorithm.  The main contributions of this paper are as follows.

{\it 1)} In order to meet the needs of various scenarios, this paper establishes the discrete {\it Pareto-optimal}  model to minimize of the ISL and PSL on the basis of previous work \cite{Kerahroodi_17}\cite{Kerahroodi_19}. In addition, this paper considers a correlation interval of interest, which can be a single shift, a segment, or a collection of them.
Compared with the existing works, the model proposed in this paper is more flexible.

{\it 2)} For the formulated design problem with discrete constraints, this paper transform it to continuous problem based on relaxation idea. To solve it efficiently, two algorithms Penalty Box ADMM and $\ell_p$-Box ADMM are proposed.
Through theoretical analysis, it is proved that the proposed methods converge to the stationary point of the approximate problem.
If the weight of penalty term in Penalty Box ADMM is reasonably chosen, the correlation performance of the generated sequence  is a litter bit better than that of  $\ell_p$-Box ADMM.
However, $\ell_p$-Box ADMM without weight parameter is easier to implement.

{\it 3)} Different from the serial execution methods in \cite{Kerahroodi_17}--\hspace{-0.01cm}\cite{Yu_20}, the proposed algorithm can be executed distributedly leading to high efficiency.
Besides, the proposed methods utilize inherent sparse structure of the optimization problem to achieve polynomial computational complexity which is lower than existing works e.g. \cite{Kerahroodi_17}--\hspace{-0.01cm}\cite{Yu_20}.  
Simulation results involving convergence and correlation are utilized to show improved performance in comparison with the state-of-the-art methods.

The rest of the paper is organized as follows. In Section II, we formulate the binary sequence design problem to a nonconvex consensus optimization problem.
In Section III, Box ADMM framework is customized to solve the nonconvex problem. To encourage binary solution, Penalty Box ADMM and $\ell_p$-Box ADMM algorithms are proposed.
The performance analysis, including convergence and computational complexity of the proposed ADMM approach, is presented in Section IV.
Finally, Section V demonstrates the effectiveness of the proposed ADMM algorithms and the conclusions are given in Section VI.

\emph{Notation}: bold lowercase and uppercase letters denote column vectors and matrices and italics denote scalars. $\mathds{R}$ and $\mathds{C}$ denote the real field and the complex field respectively.
Other notations used through this paper are presented in Table \ref{notation}.

\begin{table}[t]\small
\caption{Notations}
\label{notation}
\centering
\begin{tabular}{l|l}
\specialrule{0.1em}{1pt}{2pt}
Notation & Description \\\specialrule{0em}{1pt}{1pt}
\hline \specialrule{0em}{1pt}{1pt}
$\mathbb{R}^{N\times M}$ & the sets of $N\times M$-dimensional real-valued matrix    \\ [3pt]
$\mathbf{x}^T$ &  the transpose of vector $\mathbf{x}$  \\ [3pt]
$\mathbf{I}_N$ &  $N\times N$-dimensional identity matrix  \\ [3pt]
$|{x}|$  & the absolute value of ${x}$ \\ [3pt]
$\|\mathbf{x}\|_2$   &  the Euclidean norm of vector $\mathbf{x}$ \\[1pt]
$\|\mathbf{x}\|_p$   & the $\ell_p$-norm of vector $\mathbf{x}$, defined as $(\sum\limits_k|\mathbf{x}_k|^p)^{\frac{1}{p}}$\\
$\|\mathbf{X}\|_F$   & the Frobenius  norm of matrix $\mathbf{X}$\\[3pt]
$\|\mathbf{x}\|_{\infty}$   & the infinite norm of vector $\mathbf{x}$\\[3pt]
$\nabla f(\cdot)$      & the gradient of function $ f(\cdot)$ \\[3pt]
$\langle \mathbf{x},\mathbf{y}\rangle$      & the dot product of $\mathbf{x}$ and $\mathbf{y}$       \\  [3pt]
$\mathop {\prod} (\mathbf{x})$ & the projection of $\mathbf{x}$   \\[3pt]
\specialrule{0.1em}{1pt}{1pt}
\end{tabular}
\end{table}

\section{Problem Formulation}\label{sec:format}
\subsection{System Model}
Let consider a binary sequence set $\mathbf{X}=[\mathbf{x}_1,\mathbf{x}_2,\cdots,\mathbf{x}_M]$, and each element in sequences $\{\mathbf{x}_i\}_{i=1}^M$ is $-1$ or $1$. The aperiodic and periodic correlation functions of sequence $\mathbf{x}_i$ and $\mathbf{x}_j$ at shift $n$ are defined as
\begin{equation}\label{correlation}
\begin{split}
r_{ij,n}^{\mathcal{AP}} \!= \!\!\sum_{k=n+1}^Nx_{i,k}x_{j,k-n},r_{ij,n}^\mathcal{P} \!= \!\!\sum_{k=n+1}^Nx_{i,k}x_{j,k-n_{mod(N)}},
\end{split}
\end{equation}
where $i,j=1, \cdots ,M; n=-N+1, \cdots ,N-1$. When $i=j$, $r_{ij,n}^{\mathcal{AP}}$ and $r_{ij,n}^{\mathcal{P}}$  represent the aperiodic/periodic auto-correlation function of $\{\mathbf{x}_i\}_{i=1}^M$. Otherwise, they are cross-correlation functions.

There are some metrics that are used to evaluate the goodness of the correlation properties of binary sequences.
The most commonly used ones are the ISL \cite{Wang_12}, PSL\cite{Lin_19}, and Peak to Average Power Radio (PAPR) \cite{Dai_09}, etc.
ISL gives the relationship among the sequence and its shift version.
It is often used to measure the synchronization performance between the received signal and all the interference signals caused by multipath in the wireless communication system.
PSL is a metric derived from the autocorrelation function which describes the relationship among the maximum of the side lobes (SL).
It means the worst case of interference caused by SL to main lobe.
Therefore binary sequence sets with small PSL value are desirable.

In this paper, we consider the optimization metric of PSL.
The definitions is given in \eqref{norm metric}.
\begin{subequations}\label{norm metric}
\begin{align}
PSL:\sum_{i=1}^M\max_{l\in\mathds{L}\backslash 0}\bigg\{|r_{ii,l}|\bigg\}+\sum_{i=1}^M\sum_{\substack{j=1\\j\neq i}}^M\max_{l\in\mathds{L}}\bigg\{|r_{ij,l}|\bigg\} ,\label{norm inf}
\end{align}
\end{subequations}
where $r_{ij,l}=r_{ij,l}^{\mathcal{AP}}$ and $r_{ij,l}=r_{ij,l}^\mathcal{P}$ address the aperiodic/periodic function respectively. $\mathds{L}$ denotes the shift set interval of interest.
To facilitate the subsequent expression, we denote it as $f(\mathbf{X})$.
By utilizing off-line matrices $\mathbf{S}_l^{\mathcal{AP}}$ and cyclic shift matrices $\mathbf{S}_l^\mathcal{P}$ defined in \eqref{def Sap}, \eqref{def Sp} and denote $\mathbf{x}_i = \mathbf{X}\mathbf{s}_i$, where $i$-th element in $\mathbf{s}_i$ is $1$ and the rest elements are zeros.
\begin{equation}\label{def Sap}
\begin{split}
 &\qquad\qquad \quad l~{\rm zeros}\\
&{\bf{S}}_l^{\mathcal{AP}}=
\left[
  \begin{array}{ccccc}
     \overbrace{0~ \cdots ~0} & 1 & ~ & \mathbf{\scalebox{1.5}0}  \\
     ~ & ~ &  \ddots &~   \\
     ~ & ~ & ~ & 1  \\
     \mathbf{\scalebox{2.5}0} & ~ & ~ & ~   \\
  \end{array}
\right],
\end{split}
\end{equation}
\begin{equation}\label{def Sp}
\begin{split}
 &\qquad\qquad \quad l~{\rm zeros}\\
&{\bf{S}}_l^\mathcal{P}=
\left[
  \begin{array}{ccccc}
     \overbrace{0~ \cdots ~0} & 1 & ~ & \mathbf{\scalebox{1.5}0}  \\
     ~ & ~ &  \ddots &~   \\
     ~ & ~ & ~ & 1  \\
     \mathbf{\scalebox{2.5}I}_{l} & ~ & ~ & ~   \\
  \end{array}
\right],
\end{split}
\end{equation}
$f(\mathbf{X})$ can be rewritten as
\begin{equation}\label{ISL PSL}
\begin{split}
&f(\mathbf{X})\! = \!\!\sum_{i=1}^M\max_{l\in\mathds{L}\backslash 0}\bigg\{|(\mathbf{X}\mathbf{s}_i)^T\mathbf{S}_l\mathbf{X}\mathbf{s}_i|\bigg\}\!\!\\
&\hspace{3cm}+\!\!\sum_{i=1}^M\sum_{\substack{j=1\\j\neq i}}^M\max_{l\in\mathds{L}}\bigg\{|(\mathbf{X}\mathbf{s}_i)^T\mathbf{S}_l\mathbf{X}\mathbf{s}_j|\bigg\}\\
&\hspace{0.75cm}= \!\!\sum_{i=1}^M\sum_{j=1}^M\max_{l\in\mathds{L}}\bigg\{f_{ij,l}(\mathbf{X})\bigg\},
\end{split}
\end{equation}
where
\begin{equation}\label{def fl}
\begin{split}
f_{ij,l}(\mathbf{X})\triangleq |(\mathbf{X}\mathbf{s}_i)^T\mathbf{S}_l\mathbf{X}\mathbf{s}_j -N\delta_{|i-j|+|l|}|.
\end{split}
\end{equation}
 {In \eqref{def fl}, $\delta_{|i-j|+|l|}$ in \eqref{ISL PSL} denotes the Dirac-$\delta$ function and $\mathbf{S}_l=\mathbf{S}_l^{\mathcal{AP}}$ and $\mathbf{S}_l=\mathbf{S}_l^{\mathcal{P}}$ are aperiodic and periodic cases respectively}.
 To facilitate the subsequent expression, we define the notation set $\mathbb{O} = \{ij,l|,i,j=1,\cdots,M,l\in\mathds{L}\}$.
Then, the problem for designing  binary sequence set with good PSL can be formulated as
\begin{equation}\label{ori model}
\begin{split}
&\hspace{0.2cm}\min_{\mathbf{X}\in \mathbb{R}^{N\times M}} \hspace{0.5cm}\sum_{i,j=1}^M\max_{l\in\mathds{L}}\big\{f_{o}(\mathbf{X})\big\},\\
&{\rm  subject\ to}  \hspace{0.5cm} \mathbf{X}\in\mathcal{X},
\end{split}
\end{equation}
where
\[
\mathcal{X} = \{\mathbf{X}|x_{i,n}\in \{-1,1\},\ i=1,\cdots,M,\ n=1,\cdots,N\}.
\]

Since the constraint $\mathbf{X}\in\mathcal{X}$ is discrete, problem \eqref{ori model} is a combination optimization problem with respect to variable $\pm 1$.
That means the computational complexity of obtaining global optimal solution to problem \eqref{ori model} grows exponentially with the size of set $\mathcal{X}$.
Such a high complexity in practical application is unbearable.
The usual way to solve this problem is to relax the binary constraint to continuous box constraints, i.e., $\mathbf{X}\in\mathcal{X}_{B}$,
where
\[
\mathcal{X}_B = \left\{\mathbf{X}{\big |}\  \mathbf{x}_{i}\in[-1, 1],\ i=1,\cdots,M \right\}.
\]

The relaxed operation of binary constraint may lead the elements in $\mathbf{X}$ to  be non-binary solution during the iteration process.
To encourage binary solutions, some  {penalty method, e.g., adding the penalty term in the objective function can be introduced.
However, the non-convexity of the added penalty term may lead to further issues, namely undesirable local minima and sensitivity to the initialization.}
To tackle it, we relax the binary constraint to $\ell_p$-Box intersection.

\emph{Proposition:} $\ell_p$-Box Intersection \cite{Wu_19}: The binary set $\{-1,1\}^{N\times M}$ can be equivalently replaced by the intersection between a sphere $\mathcal{X}_S$ and a box $\mathcal{X}_B$, as follows:
\begin{equation}\label{def lp box}
\begin{split}
\mathbf{X}\!\in\! \{-1,1\}^{N\!\times\! M} \!\Leftrightarrow \!\mathbf{X}\!\in\! \mathcal{X}_S \bigcap\mathcal{X}_B.
\end{split}
\end{equation}
\begin{figure}[t]
  \centering
  \centerline{\includegraphics[scale=0.6]{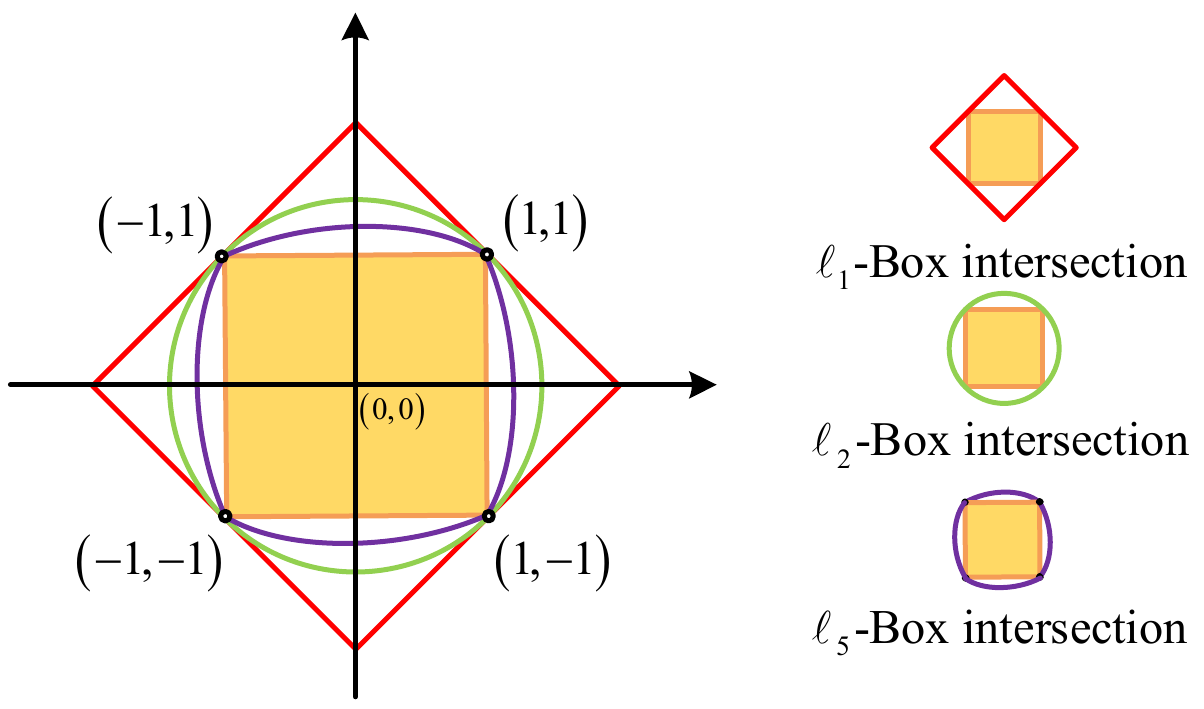}}
\caption{Geometric illustration of the equivalence between $\ell_p$-Box intersection and the set of binary points in $\mathbb{R}^2$, $p=\{1,2,5\}$.}
\label{fig lp box}
\end{figure}
where
\begin{equation}\label{def xs}
\begin{split}
\mathcal{X}_S = \left\{ \mathbf{X}:\!\|\mathbf{X}\|_p^p \!=\!MN \right\}.
\end{split}
\end{equation}
Note that $\mathcal{X}_S$ can be seen as a $(MN- 1)$-dimensional $\ell_p$-sphere centered at origin of axes with radius $(MN)^{\frac{1}{p}}$ and $p\in(0,\infty)$.
To illustrate  this proposition, we present a $2$-dimensional example with different $p$ parameters in Fig. \ref{fig lp box}.
It is obvious from the figure that binary set $\{-1,1\}^{2}$ is the intersection between  the sphere $\mathcal{X}_S$ and the box $\mathcal{X}_B$.

By relaxing the binary constraint to $\ell_p$-Box and introducing auxiliary variables $\{\mathbf{Z}_o\in \mathbb{R}^{N\times M}\}_{o\in \mathbb{O}}$, problem \eqref{ori model} can be transformed into the following global consensus problem
\begin{equation}\label{model consensus}
\begin{split}
 \hspace{-0.2cm}&\min_{\{\mathbf{X}_1,\mathbf{X}_2,\mathbf{Z}_o\}\in \mathbb{R}^{N\times M}} \sum_{i,j=1}^M\max_{l\in\mathds{L}}\big\{f_o(\mathbf{Z}_o)\big\},\\
 \hspace{-0.2cm}&{\rm  subject\ to} \ \mathbf{X}_1\!\in\!\mathcal{X}_B, \!\mathbf{X}_2\!\in\!\mathcal{X}_S,\! \mathbf{X}_1 \!=\! \mathbf{Z}_o,  \mathbf{X}_2\! =\!
 \mathbf{Z}_o,o\in \mathbb{O}.
\end{split}
\end{equation}
The introduction of auxiliary variables makes each subproblem have its local variables. Thus, each subproblem can be solved in parallel.
Define the following matrices
\begin{equation}\label{def lp new}
\begin{split}
&\bar{\mathbf{X}} = [\mathbf{X}_1;\mathbf{X}_2], \ \mathbf{A} = [\mathbf{I}_N;\mathbf{I}_N].
\end{split}
\end{equation}
The linear constraints can be rewritten as $\mathbf{A}\mathbf{Z}_o=\bar{\mathbf{X}} $.

\section{Solving Algorithm}
In this section, one algorithm named $l_p$-Box ADMM algorithm is developed to solve problem \eqref{model consensus}.
In comparison with the state-of-the-art methods, one major benefit of proposed algorithm is the parallel execution structure.
Another benefit is that the feasible region of the optimization variable is relaxed to the intersection of two continuous regions.
The former one benefit can greatly improve the algorithm execution's efficiency.
The latter one can promote the sequence with better correlation performance.

\subsection{$\ell_p$-Box ADMM Algorithm}

 In \eqref{def lp new}, $\mathbf{A}^T=[\mathbf{I}_N,\mathbf{I}_N]\in\mathbb{R}^{N\times2N}$ is full row rank, i.e., $rank(\mathbf{A}^T)=N<2N$.
Main difficulty during the convergence analysis is the constraint $ \mathbf{A}\mathbf{Z}_o =\mathbf{X}$.
To tackle it, define
$ \bar{\mathbf{A}} = [\mathbf{A},\sigma\mathbf{I}_{2N}]\in\mathbb{R}^{2N\times3N},\ \bar{\mathbf{Z}}_o = [\mathbf{Z}_o; \hat{\mathbf{Z}}_{o}],$
where $[\hat{\mathbf{Z}}_{o1};\hat{\mathbf{Z}}_{o2}]=\hat{\mathbf{Z}}_{o}$ are introduced perturbed variable to construct constraint $\bar{\mathbf{A}}\bar{\mathbf{Z}}_o=\bar{\mathbf{X}} $, i.e.,
\begin{equation}\label{def cons}
\begin{split}
\mathbf{A}\mathbf{Z}_o+\sigma\hat{\mathbf{Z}}_{o}=\bar{\mathbf{X}} ,
\end{split}
\end{equation}
where $\sigma>0$ is sufficiently small.
Then, the associated perturbed problem is presented as
\begin{equation}\label{new lp model}
\begin{split}
 \hspace{-0.2cm}&\min_{\{\mathbf{X},\bar{\mathbf{Z}}_o\}\in \mathbb{R}^{N\times M}} \sum_{i,j=1}^M\left(\max_{l\in\mathds{L}}\big\{f_o(\mathbf{Z}_o)\big\}+ \frac{\sigma^2}{2}\hat{\mathbf{Z}}_{o}^{T}\hat{\mathbf{Z}}_{o}\right),\\
 \hspace{-0.2cm}&{\rm  subject\ to} \ \mathbf{X}_1\!\in\!\mathcal{X}_B, \mathbf{X}_2\!\in\!\mathcal{X}_S, \bar{\mathbf{A}}\bar{\mathbf{Z}}_o=\bar{\mathbf{X}} , o\in \mathbb{O}.
\end{split}
\end{equation}

  {ADMM, a popular technique, is suitable for solving problem \eqref{model consensus} with multiple separable subproblems.}
The corresponding augmented Lagrangian function can be expressed as
\begin{equation}\label{lag penalty}
\begin{split}
&\mathcal{L} \left(\bar{\mathbf{X}} ,\{ \bar{\mathbf{Z}}_o,\mathbf{\Lambda }_o\}_{o \in \mathbb{O}}\right) \\
  = \ &\sum_{i,j=1}^M\left(\max_{l\in\mathds{L}}\big\{f_o(\mathbf{Z}_o)\big\}+ \frac{\sigma^2}{2}\hat{\mathbf{Z}}_{o}^{T}\hat{\mathbf{Z}}_{o}\right)\\
  &\hspace{0.6cm}{ + \sum\limits_{o \in \mathbb{O}} {\left( {\left\langle {\mathbf{\Lambda }}_o,\bar{\mathbf{X}}  - \bar{\mathbf{A}}\bar{\mathbf{Z}}_o \right\rangle  + \frac{{{\rho_o}}}{2}\left\| {\bar{\mathbf{X}}  - {\bar{\mathbf{A}}\bar{\mathbf{Z}}_o}} \right\|_F^2} \right)} }.
\end{split}
\end{equation}
where $\{\mathbf{\Lambda}_o\in \mathbb{R}^{N\times M},\rho_o>0\}_{o \in \mathbb{O}}$ are Lagrangian multipliers and penalty parameters respectively.
Thus, the $\ell_p$-Box ADMM framework to solve problem \eqref{model consensus} can be described as
\begin{subequations}\label{lp ADMM}
\begin{align}
&\bar{\mathbf{X}} ^{k+1} = \underset{\mathbf{X}_1\in\mathcal{X}_B,\mathbf{X}_2\in\mathcal{X}_S} {\arg \min}\ \ \mathcal{L}\left(\bar{\mathbf{X}} ,\{ \bar{\mathbf{Z}}_o^k,\mathbf{\Lambda }_o^k\}_{o \in \mathbb{O}}\right),\label{s1 lp ADMM}\\
&\bar{\mathbf{Z}}_o^{k+1} =  \underset{\bar{\mathbf{Z}}_o} {\arg \min}\ \ \mathcal{L}\left(\bar{\mathbf{X}} ^{k+1}, \bar{\mathbf{Z}}_o,\mathbf{\Lambda }_o^k\right),o \in \mathbb{O},\label{s2 lp ADMM}\\
&\mathbf{\Lambda}_o^{k+1} = \mathbf{\Lambda}_o^{k} + \rho_o(\bar{\mathbf{X}} ^{k+1} -\bar{\mathbf{A}}\bar{\mathbf{Z}}_o^{k+1}),o \in \mathbb{O},\label{s3 lp ADMM}
\end{align}
\end{subequations}
where $k$ denotes the iteration number.
 {Since function $\mathcal{L}(\cdot)$ is nonconvex related to variable $\mathbf{Z}_o$.
The challenge of implementing Penalty Box ADMM is how to solve \eqref{s2 lp ADMM}.}

\subsubsection{Solving Subproblem \eqref{s1 lp ADMM}}
Problem \eqref{s1 lp ADMM}\ can be equivalent to the following problem
\begin{equation}\label{equ pro s1}
\begin{split}
 \hspace{-0.3cm}&\min_{\bar{\mathbf{X}} \in \mathbb{R}^{N\!\times\! M}} \sum\limits_{o \in \mathbb{O}} {\left( {\left\langle {\mathbf{\Lambda }}_o^k,\bar{\mathbf{X}}  - \bar{\mathbf{A}}\bar{\mathbf{Z}}_o^k \right\rangle  + \frac{{{\rho_o}}}{2}\left\| {\bar{\mathbf{X}}  - {\bar{\mathbf{A}}\bar{\mathbf{Z}}_o^k}} \right\|_F^2} \right)} ,\\
 \hspace{-0.3cm}&\ {\rm  subject\ to}  \hspace{0.2cm}\ \mathbf{X}_1\in\mathcal{X}_B,\mathbf{X}_2\in\mathcal{X}_S.
\end{split}
\end{equation}
In problem \eqref{equ pro s1}, objective function is  quadratic w.r.t. $\bar{\mathbf{X}} $ and  $\mathbf{X}_1\in\mathcal{X}_B,\mathbf{X}_2\in\mathcal{X}_S$.
Thus, the optimal solution to \eqref{equ pro s1} can be obtained through the following procedures: set the gradient function to be zero, i.e.,
\begin{equation}\label{grad x}
\begin{split}
 \nabla_{\bar{\mathbf{X}} } {\left[ \sum\limits_{o \in \mathbb{O}} {\left( {\left\langle {{\mathbf{\Lambda }}_o^k,{\bar{\mathbf{X}} }\! - \! \bar{\mathbf{A}}\bar{\mathbf{Z}}_o^k} \right\rangle  + \frac{{{\rho_o}}}{2}\left\| {{\bar{\mathbf{X}} } \!- \! \bar{\mathbf{A}}\bar{\mathbf{Z}}_o^k} \right\|_F^2} \right)}\right]}\!=\!0.
\end{split}
\end{equation}
Then, we obtain its solution
\begin{equation}\label{solu x}
\begin{split}
 \hat{\mathbf{X}}^{k+1} =  \left[
 \begin{array}{c}
     \hat{\mathbf{X}}_1^{k+1}\\
  \hat{\mathbf{X}}_2^{k+1} \\
  \end{array}
 \right]=\frac{{\mathop \sum \limits_{o \in \mathbb{O}} ({\rho _o} \bar{\mathbf{A}}\bar{\mathbf{Z}}_o^k - {\mathbf{\Lambda }}_o^k)}}{{\mathop \sum \limits_{o \in \mathbb{O}} \rho _o}}.
\end{split}
\end{equation}
Projecting $ \hat{\mathbf{X}}_1^{k+1}$ and $ \hat{\mathbf{X}}_2^{k+1}$ onto $\mathcal{X}_B$ and $\mathcal{X}_S$ respectively, we can obtain
\begin{equation}\label{solution x}
\begin{split}
{\mathbf{X}}_1^{k+1} = \mathop \prod \limits_{\mathbf{X}\in\mathcal{X}_B} \left( \hat{\mathbf{X}}_1^{k+1} \right), {\mathbf{X}}_2^{k+1} = \mathop \prod \limits_{\mathbf{X}\in\mathcal{X}_S} \left( \hat{\mathbf{X}}_2^{k+1} \right),
\end{split}
\end{equation}
where $\mathop {\prod} \limits_{\mathbf{X}\in \mathcal{X}_B}(\cdot)$ project every entry of the input variable onto $[-1,1]$ and
\begin{equation}\label{def pro xs}
\begin{split}
\mathop \prod \limits_{\mathbf{X}\in\mathcal{X}_S} \left(\mathbf{X}\right)= \frac{ \mathbf{X}}{\|\mathbf{X}\|_p}(MN)^{1/p}.
\end{split}
\end{equation}
Problem \eqref{s1 lp ADMM}'s solution is $\bar{\mathbf{X}}^{k+1} =[{\mathbf{X}}_1^{k+1};{\mathbf{X}}_2^{k+1}]$.

\subsubsection{Solving Subproblem \eqref{s2 lp ADMM}}
Since \eqref{s2 lp ADMM} is an unconstrained problem, the major challenge to solve it is how to handle $\mathop {\max }\limits_{l\in\mathds{L}} \left\{ f_o(\mathbf{Z}_o) \right\}$.
Define the following vector
\[
\begin{split}
\mathbf{f}_{ij}(\{\mathbf{Z}_o\}) = \left[\cdots,f_{ij,l}(\mathbf{Z}_{ij,l}),f_{ij,l+1}(\mathbf{Z}_{ij,l+1}),\cdots\right]\in\mathbb{R}^{|\mathds{L}|}.
\end{split}
\]
Then, $\mathop {\max }\limits_{l\in\mathds{L}} \left\{ f_o(\mathbf{Z}_o) \right\}$ is equivalent to $\|\mathbf{f}_{ij}(\{\mathbf{Z}_o\}) \|_{\infty}$.
Given that from an analytical point of view, the $\ell_{\infty}$-norm is not a well-behaved function, $\ell_q$-norms will be used instead, i.e.,
\begin{equation}\label{psl p norm}
\begin{split}
\hspace{-0.2cm} \mathop {\max }\limits_{l\in\mathds{L}} \left\{ f_o(\mathbf{Z}_o) \right\} \!=\! \lim_{q\rightarrow\infty}\|\mathbf{f}_{ij}(\{\mathbf{Z}_o\})\|_{q}\! = \! \left(\sum_{l\in\mathds{L}}f_o^q(\mathbf{Z}_o)\right)^{\!\frac{1}{q}},
\end{split}
\end{equation}
where $q>2$ is a integer.
Minimizing $\left(\sum\limits_{l\in\mathds{L}}f_o^q(\mathbf{Z}_o)\right)^{\!\frac{1}{q}}$ is equivalent to minimizing $\sum\limits_{l\in\mathds{L}}f_o^q(\mathbf{Z}_o)$.
Assuming the current iteration index is $k$, $f_o^q(\mathbf{Z}_o)$ can be majored at point $\mathbf{Z}_o^{k}$ \cite{Song_16a} \cite{Fan_19}, i.e.,
\begin{equation}\label{clp}
\begin{split}
f_o^q(\mathbf{Z}_o)\leq \frac{q}{2}f_o^{q-2}(\mathbf{Z}_o^k)f_o^2(\mathbf{Z}_o) + f_{cons}.
\end{split}
\end{equation}
where $f_{cons}$ is the constant term.
Plugging \eqref{psl p norm} and \eqref{clp} into problem \eqref{s2 lp ADMM} and dropping the constant term, we get  the
following approximate problem
\begin{equation}\label{equ pro s2}
\begin{split}
 \hspace{-0.3cm} \min_{\{\bar{\mathbf{Z}}_o\}}   \mathcal{L} \left(\bar{\mathbf{X}},\{ \bar{\mathbf{Z}}_o,\mathbf{\Lambda }_o\}_{o \in \mathbb{O}}\right),
\end{split}
\end{equation}
where
\begin{equation}\label{app lag}
\begin{split}
&\mathcal{L} \left(\bar{\mathbf{X}},\{ \bar{\mathbf{Z}}_o,\mathbf{\Lambda }_o\}_{o \in \mathbb{O}}\right) \\
  = \ &\sum_{o \in \mathbb{O}} \left(w_of_o^2(\mathbf{Z}_o) + \sigma^2\hat{\mathbf{Z}}_{o}^{T}\hat{\mathbf{Z}}_{o}\right.\\
  &\hspace{1.4cm}  \left. + {\left\langle {\mathbf{\Lambda }}_o,{\bar{\mathbf{X}}} - \bar{\mathbf{A}}\bar{\mathbf{Z}}_o \right\rangle  + \frac{{{\rho_o}}}{2}\left\| {{\bar{\mathbf{X}}} - {\bar{\mathbf{A}}\bar{\mathbf{Z}}_o}} \right\|_F^2} \right) \\
  = \ & \sum_{o \in \mathbb{O}}\mathcal{L}_o \left(\bar{\mathbf{X}}, \bar{\mathbf{Z}}_o,\mathbf{\Lambda }_o\right).
\end{split}
\end{equation}
where $w_o$ is the normalized weight
\begin{equation}\label{def lp}
\begin{split}
{w}_o = \left(\frac{f_o(\mathbf{Z}_o^k)}{\mathop {\max }\limits_{o \in \mathbb{O}} \left\{ f_o(\mathbf{Z}_o^k) \right\}}\right)^{q-2}.
\end{split}
\end{equation}

For all $o \in \mathbb{O}$, subproblems in \eqref{equ pro s2} are independent of each other and each subproblem is unconstrained w.r.t. variable $\mathbf{Z}_o$.
That means subproblems in \eqref{equ pro s2} can be implemented in parallel.
However, solving problem \eqref{equ pro s2} is still difficult, since $f_o^2(\mathbf{Z}_o)$ is nonconvex related to $\mathbf{Z}_o$ (see \eqref{def fl}).
To tackle it, we have the following lemma, which indicates that $\{f_o^2(\mathbf{Z})\}_{o \in \mathbb{O}}$ are continuous, differentiable and have Lipschitz continuous gradients in the finite domain $\mathcal{X}_c= \left\{\mathbf{X}{\big |}\  \mathbf{{x}}_{i}\in[-c, c],\ i=1,\cdots,M \right\}$ of the point $\mathbf{\hat{X}}\in \mathcal{X}_{B}$ (see proof in Appendix \ref{proof lemma lip}).
 {
\begin{lemma}\label{Lips cont}
gradients $\nabla f_o(\mathbf{X})$ are Lipschitz continuous, i.e.,
\begin{equation}\label{Lipschitiz}
\begin{split}
 \|\nabla f^2_o(\mathbf{\tilde{X}})\!-\!\nabla f_o^2(\mathbf{\hat{X}})\|_F\! \leq\! L_o\|\mathbf{\tilde{X}}\!-\!\mathbf{\hat{X}}\|_F, \ o \in \mathbb{O},
\end{split}
\end{equation}
where $\mathbf{\tilde{X}} \in \mathcal{X}_c$, $\mathbf{\hat{X}}\in\mathcal{X}_B$  and constants
\begin{equation}\label{Lips cons}
\begin{split}
L_o\!\geq\! 2(N+1){\hat c}^2, \ {\hat c} = \max\{c,1\}.
\end{split}
\end{equation}
\end{lemma}}
Based on Lemma \ref{Lips cont} and Decent Lemma in \cite{Bertsekas_99}, we have
\begin{equation}\label{upp bound fun}
\begin{split}
&\mathcal{L}_{o}\left(\bar{\mathbf{X}}^{k+1}, \bar{\mathbf{Z}}_o,\mathbf{\Lambda }_o^k\right)\\
 \leq & \  {w}_o f_o^2(\mathbf{X}^{k+1}) + \!\!\left\langle {w}_o\nabla f^2_o(\mathbf{X}^{k+1}),\mathbf{Z}_o-\mathbf{X}^{k+1} \right\rangle \\
&\hspace{0.1cm}+ \frac{{w}_oL_o}{2}\left\| {\mathbf{X}^{k+1}-{\mathbf{Z}}_o} \right\|_{F}^{2} + \left\langle \mathbf{\Lambda }_o^k,\bar{\mathbf{X}}^{k+1} -\bar{\mathbf{A}}\bar{\mathbf{Z}}_o\right\rangle\\
&\hspace{0.1cm} +\frac{\rho_o}{2}\left\| {\bar{\mathbf{X}}^{k+1}-\bar{\mathbf{A}}\bar{\mathbf{Z}}_o} \right\|_{F}^{2} + \frac{\sigma^2}{2}\hat{\mathbf{Z}}_{o}^{T}\hat{\mathbf{Z}}_{o}.
\end{split}
\end{equation}
Define the right-hand side of the above inequality as $\mathcal{U}_{o}(\mathbf{Z}_o)$.
We customize the $\ell_p$-Box ADMM by minimizing it instead of \eqref{equ pro s2}.
Since $\mathcal{U}_{o}(\mathbf{Z}_o)$ is convex quadratic function w.r.t. $\mathbf{Z}_o$, the optimal solution can be obtained by setting $\nabla_{\mathbf{Z}_o}\mathcal{U}_{o}(\mathbf{Z}_o) = 0$.
Through solving the equation, we get the solution
\begin{subequations}\label{solu z}
\begin{align}
\hspace{-0.2cm}\hat{\mathbf{Z}}_o^{k+1}& = \frac{\mathbf{\Lambda}_o^k+\rho_o(\bar{\mathbf{X}}^{k+1} - \mathbf{A}\mathbf{Z}_o^{k})}{(\rho_o+1)\sigma},
\\
\hspace{-0.2cm}\mathbf{Z}_o^{k+1} &\!= \!\mathbf{X}^{k+1}\! + \!\frac{\mathbf{A}^T(\mathbf{\Lambda}_o^k \!- \!\rho_o\sigma\hat{\mathbf{Z}}_o^{k+1})\!-\! {w}_o\nabla f^2_o(\mathbf{X}^{k+1})}{2\rho_o+{w}_oL_o},
\end{align}
\end{subequations}
where
\begin{equation}\label{def xo}
\begin{split}
\mathbf{X}^{k+1} = \frac{\mathbf{A}^T(\bar{\mathbf{X}}^{k+1}-\sigma\hat{\mathbf{Z}}_o^{k+1})}{2}.
\end{split}
\end{equation}

Combining \eqref{s3 lp ADMM}, \eqref{solution x} and \eqref{solu z}, we summarize the customized Penalty Box ADMM algorithm in Table \ref{table lp Box ADMM}.

\begin{table}[tbp]
\renewcommand \arraystretch{1.2}
\caption{The customized Penalty Box ADMM algorithm }
\label{table lp Box ADMM}
\centering
\begin{tabular}{l}
 \hline\hline
 {\bf Initialization:} Compute Lipschitz constants $\{L_o,\!l\!\in\!\mathds{L}\}$  \\
 \hspace{0.2cm} according to \eqref{Lips cons}.
  Set iteration index $k\!=\!1$, initialize \\
 \hspace{0.2cm} $\bar{\mathbf{X}}^1$ and $\{\mathbf\Lambda_o^1, o \in \mathbb{O}\}$ randomly, and let $\{\bar{\mathbf{X}}^1= \bar{\mathbf{A}}\bar{\mathbf{Z}}_o^1,$ \\
 \hspace{0.2cm} $ o \in \mathbb{O}\}$. \\
  {\bf repeat} \\
  \hspace{0.2cm} S.1 Compute ${\bar{\mathbf{X}}}^{k+1} $ via \eqref{solu x} and \eqref{solution x} , i.e., \\[5pt]
  \hspace{0.9cm} $ \bar{\mathbf{X}}^{k+1} =  \left[
 \begin{array}{c}
      \mathop \prod \limits_{\mathbf{X}\in\mathcal{X}_B} \left( \hat{\mathbf{X}}_1^{k+1} \right)\\
 \mathop \prod \limits_{\mathbf{X}\in\mathcal{X}_S} \left( \hat{\mathbf{X}}_2^{k+1} \right) \\
  \end{array}
 \right].$\\[20pt]
  \hspace{0.2cm} S.2 Compute $\{\bar{\mathbf{Z}}_o^{k+1},o \in \mathbb{O}\}$ via \eqref{solu z} in parallel, i.e.,\\[5pt]
  \hspace{0.9cm}
  $
   \hat{\mathbf{Z}}_o^{k+1} = \frac{\mathbf{\Lambda}_o^k+\rho_o(\bar{\mathbf{X}}^{k+1} - \mathbf{A}\mathbf{Z}_o^{k})}{(\rho_o+1)\sigma}$,\\[8pt]
\hspace{1.0cm}$\mathbf{Z}_o^{k+1} = \mathbf{X}^{k+1}\! + \!\frac{\mathbf{A}^T(\mathbf{\Lambda}_o^k \!- \!\rho_o\sigma\hat{\mathbf{Z}}_o^{k+1})\!-\! {w}_o\nabla f^2_o(\mathbf{X}^{k+1})}{2\rho_o+{w}_oL_o}.
  $ \\[5pt]
  \hspace{0.2cm} S.3 Compute $\{\mathbf{\Lambda}_o^{k+1}, o \in \mathbb{O}\}$ via \eqref{s3 lp ADMM} in parallel, i.e., \\[5pt]
  \hspace{0.9cm} $\mathbf{\Lambda}_o^{k+1}\! = \mathbf{\Lambda}_o^{k} + \rho_o(\bar{\mathbf{X}} ^{k+1} -\bar{\mathbf{A}}\bar{\mathbf{Z}}_o^{k+1})$.\\
 {\bf until} some preset termination criterion is satisfied.\\
 \hspace{0.75cm} Let $\mathbf{X}^{k+1}$ be the output.\\
 \hline\hline
\end{tabular}
\end{table}

\emph{Remarks on proposed ADMM algorithms:}\\
$\bullet$ In model \eqref{model consensus}, we relax the problem to continuous variable problem. Different from traditional methods  {\cite{Kerahroodi_17}\cite{Kerahroodi_19}}, the proposed algorithms optimize variable $\mathbf{X}$ as a whole rather than individual elements of $\mathbf{X}$. \\
$\bullet$ During the iteration of the proposed algorithms, we introduce the convex quadratic function $\mathcal{L}(\cdot)$. Even if the optimization and projection operation are performed separately,  both $\prod \limits_{\mathbf{X}\in\mathcal{X}_B}(\cdot)$ and $\mathop \prod \limits_{\mathbf{X}\in\mathcal{X}_S}(\cdot)$ can guarantee the obtained solution is (locally) optimal.\\
$\bullet$ Parallel execution architecture of the proposed algorithms plays an essential role leading to better implementation efficiency than  the state-of-the-art methods.\\
$\bullet$ The proposed algorithms are guaranteed convergent to some stationary point of the nonconvex optimization problem if proper parameters are chosen. We have  {several theorems} in the following section.

\section{Algorithm Analysis}
In this section,  we show several analyses on the proposed ADMM algorithm, such as convergence and computational complexity.
The proposed algorithms are convergent to some stationary point of the approximate problem.
To reduce the algorithm's computational complexity, we exploit the inherent sparsity characteristic of the problem.

\subsection{Convergence Issue}
Before presenting the convergence conclusion, we first give a few corresponding lemmas.

\begin{lemma}\label{successive difference}
For the proposed algorithm, the augmented Lagrangian function has the following inequality
\begin{equation}\label{succ-diff phi}
\begin{split}
\hspace{-0.3cm}&  \mathcal{L}\left(\bar{\mathbf{X}}^{k},\{ \bar{\mathbf{Z}}_o^{k},\mathbf{\Lambda }_o^{k}\}_{o \in \mathbb{O}}\right)\!-\!\mathcal{L}\left(\bar{\mathbf{X}}^{k+1},
\{ \bar{\mathbf{Z}}_o^{k+1},\mathbf{\Lambda }_o^{k+1}\}_{o \in \mathbb{O}}\right)\\
\hspace{-0.3cm}&\geq \!\sum\limits_{o \in \mathbb{O}} \frac{1}{{2\rho _o^2}} \left(\bar{\epsilon}_o\left\| {{\bar{\mathbf{X}}^{k + 1}} - {\bar{\mathbf{X}}^k}} \right\|_F^2 + {\epsilon}_o \left\| {{\mathbf{Z}}_o^{k + 1} - {\mathbf{Z}}_o^k} \right\|_F^2\right.\\
&\hspace{4.5cm}\left. + \hat{\epsilon}_o \left\| {\hat{\mathbf{Z}}_o^{k + 1} - \hat{\mathbf{Z}}_o^k} \right\|_F^2\right).
\end{split}
\end{equation}
In each $\ell_p$-Box ADMM iteration, if ${\epsilon}_{o},\bar{\epsilon}_{o},\hat{\epsilon}_{o}\geq0$, $\mathcal{L}\left(\bar{\mathbf{X}}^{k},\{ \bar{\mathbf{Z}}_o^{k},\mathbf{\Lambda }_o^{k}\}_{o \in \mathbb{O}}\right)$ {\it decreases sufficiently}.
\end{lemma}

\begin{lemma}\label{lemma lower bound}
 If $\rho_n\geq5L_n$, the augmented Lagrangian function is lower bounded, i.e.,
 \begin{equation}\label{Lgeq0}
 \mathcal{L}\left(\bar{\mathbf{X}}^{k},\{ \bar{\mathbf{Z}}_o^{k},\mathbf{\Lambda }_o^{k}\}_{o \in \mathbb{O}}\right)\geq 0 , \forall k.
 \end{equation}
 \end{lemma}
The proof of Lemma \ref{successive difference} and \ref{lemma lower bound} is given in Appendix \ref{lemma converg}.

Lemma \ref{successive difference} and \ref{lemma lower bound} show that augmented Lagrangian function $\mathcal{L}(\cdot)$ decreases sufficiently and has a lower bound, which indicates it is convergent.
We have the following theorem to characterize the proposed $\ell_p$-Box ADMM algorithm.
\begin{theorem}\label{theorem converg}
$\forall {o \in \mathbb{O}}$ if penalty parameters $\rho_o$ and  Lipschitz  constants $L_o$ satisfy some wild conditions, the proposed ADMM algorithms converge to some stationary point $\mathbf{X}^*$ of approximate problem \eqref{def Fx}, i.e.,
\begin{equation}\label{theorem 1}
\begin{split}
\left\langle \nabla F(\mathbf{X}^*),\mathbf{X}- \mathbf{X}^*\right\rangle \geq 0,
\end{split}
\end{equation}
where
\begin{equation}\label{def Fx}
\begin{split}
 F(\mathbf{X}) = \sum_{o \in \mathbb{O}} w_of_o^2(\mathbf{X}).
\end{split}
\end{equation}
The detail of the proof and convergence conditions are given in Appendix \ref{proof of throrem}.
\end{theorem}


\subsection{ Implementation Analysis}
Observing the proposed ADMM algorithm, the computational cost is mainly multiplication of solving the gradient $\nabla f^2_o(\mathbf{X})$.
Function $f^2_o(\mathbf{X})$ is defined as
\begin{equation}\label{cl}
\begin{split}
f^2_o(\mathbf{X})\triangleq|\mathbf{s}_i^T\mathbf{X}^T\mathbf{S}_l\mathbf{X}\mathbf{s}_j -N\delta_{|i-j|+|l|}|^2.
\end{split}
\end{equation}
For gradient $\nabla f^2_o(\mathbf{X})$, we have
\begin{equation}\label{grad}
\begin{split}
&\hspace{-0.3cm}\nabla f^2_o(\mathbf{X})\!\!=\!\!\!
\left[
  \begin{array}{ccccccc}
    \frac{ \partial f^2_o(\mathbf{X})}{\partial x_{1,1}} &     \cdots &  \!\!\frac{ \partial f^2_o(\mathbf{X})}{\partial x_{m,1}} \!\! & \cdots  & \frac{ \partial f^2_o(\mathbf{X})}{\partial x_{M,1}} \\
    \frac{ \partial f^2_o(\mathbf{X})}{\partial x_{1,2}}  & \cdots&\frac{ \partial f^2_o(\mathbf{X})}{\partial x_{m,2}}  & \cdots  & \frac{ \partial f^2_o(\mathbf{X})}{\partial x_{M,2}}   \\
   \vdots   & \vdots &\vdots &\ddots & \vdots  \\
   \frac{ \partial f^2_o(\mathbf{X})}{\partial x_{1,N}} &\cdots &  \!\!\frac{ \partial f^2_o(\mathbf{X})}{\partial x_{m,N}} \!\! & \cdots  & \frac{ \partial f^2_o(\mathbf{X})}{\partial x_{M,N}} \\
  \end{array}
\right]\!\!,
\end{split}
\end{equation}
where
\begin{equation}\label{partial cl}
\begin{split}
\hspace{-0.3cm}&\ \ \frac{ \partial f^2_o(\mathbf{X})}{\partial x_{i,n}}\\
\hspace{-0.2cm}& = \!2\text{Tr}\left(\!\frac{ \partial (\mathbf{s}_i^T\mathbf{X}^T\mathbf{S}_l\mathbf{X}\mathbf{s}_j)^T}{\partial x_{i,n}}(\mathbf{s}_i^T\mathbf{X}^T\mathbf{S}_l\mathbf{X}\mathbf{s}_j\!\! -\!\!N\delta_{|i-j|+|l|}) \!\right).
\end{split}
\end{equation}
For $\frac{ \partial (\mathbf{s}_i^T\mathbf{X}^T\mathbf{S}_l\mathbf{X}\mathbf{s}_j)}{\partial x_{i,n}}$,
when $l=0$, we have
\begin{equation}\label{XX}
\begin{split}
&\hspace{-0.3cm}\frac{\partial \mathbf{X}^T\mathbf{X}}{\partial x_{i,n}}\!\!=\!\!\!
\left[
  \begin{array}{ccccccc}
    ~ &   n\!\!-\!\!1&\hspace{-0.7cm}  {\rm zeros} &  \!\!{x_{i,1}} \!\!& ~ & ~ & ~ \\
    ~ &  \! \!\! \overbrace{\!\mathbf{\scalebox{3.0}0}\!}\!\!\! & ~& \vdots & ~  & \!\! \mathbf{\scalebox{3.0}0}\!\!  & ~ \\
   ~  &~ &~ & \!\!{x_{i,n-1}}\!\!&~ &~  \\
   \!\!\!{x_{i,1}} &\!\!\cdots\!\! &\!{x_{i,n-1}}\! & 0 & \!\!{x_{i,n+1}}\! \!&  \!\!\cdots\!\! & \!{x_{i,M}}\!\!\!  \\
    ~ & ~ &~ &\!\!{x_{i,n+1}} \!\! &~ & ~& ~  \\
    ~ &  \!\!\mathbf{\scalebox{3.0}0} \!\! & ~ & \vdots & ~ & \!\! \mathbf{\scalebox{3.0}0} \!\! & ~  \\
    ~ & ~ &~ & \!\!{x_{i,M}}\!\! &~ & ~& ~  \\
  \end{array}
\right]\!\!.
\end{split}
\end{equation}
From \eqref{XX}, it can be found that there are $2(M-1)$ nonzero elements in $\frac{\partial \mathbf{X}^T\mathbf{X}}{\partial x_{i,n}}$.
It implies that obtaining $\frac{ \partial f^2_o(\mathbf{X})}{\partial \mathbf{X}}$ needs $2N$ multiplications at most.
Since $\mathbb{O} = \{ij,l|,i,j=1,\cdots,M,l\in\mathds{L}\}$, there are $M^2|\mathds{L}|$ entries in set $\mathbb{O}$, obtaining all $\{\nabla f^2_o(\mathbf{X}),o \in \mathbb{O}\}$ needs $2M^2N|\mathds{L}|$ multiplications ($|\mathds{L}|$ denotes set $\mathds{L}$'s size).
Observing Table \ref{table lp Box ADMM}, we can see that the computational cost of other terms is far less than $\nabla f^2_o(\mathbf{X})$.
Therefore, we can conclude that the computational cost of the proposed ADMM algorithm is $\mathcal{O}(M^2N|\mathds{L}|)$\footnotemark.
\footnotetext{It should be noted that the proposed algorithm can be implemented in parallel, the computational time will not increase proportionally with $|\mathds{L}|$.}

\section{Simulation Results}


\begin{figure}[t]
\centering

\centerline{\includegraphics[scale=0.6]{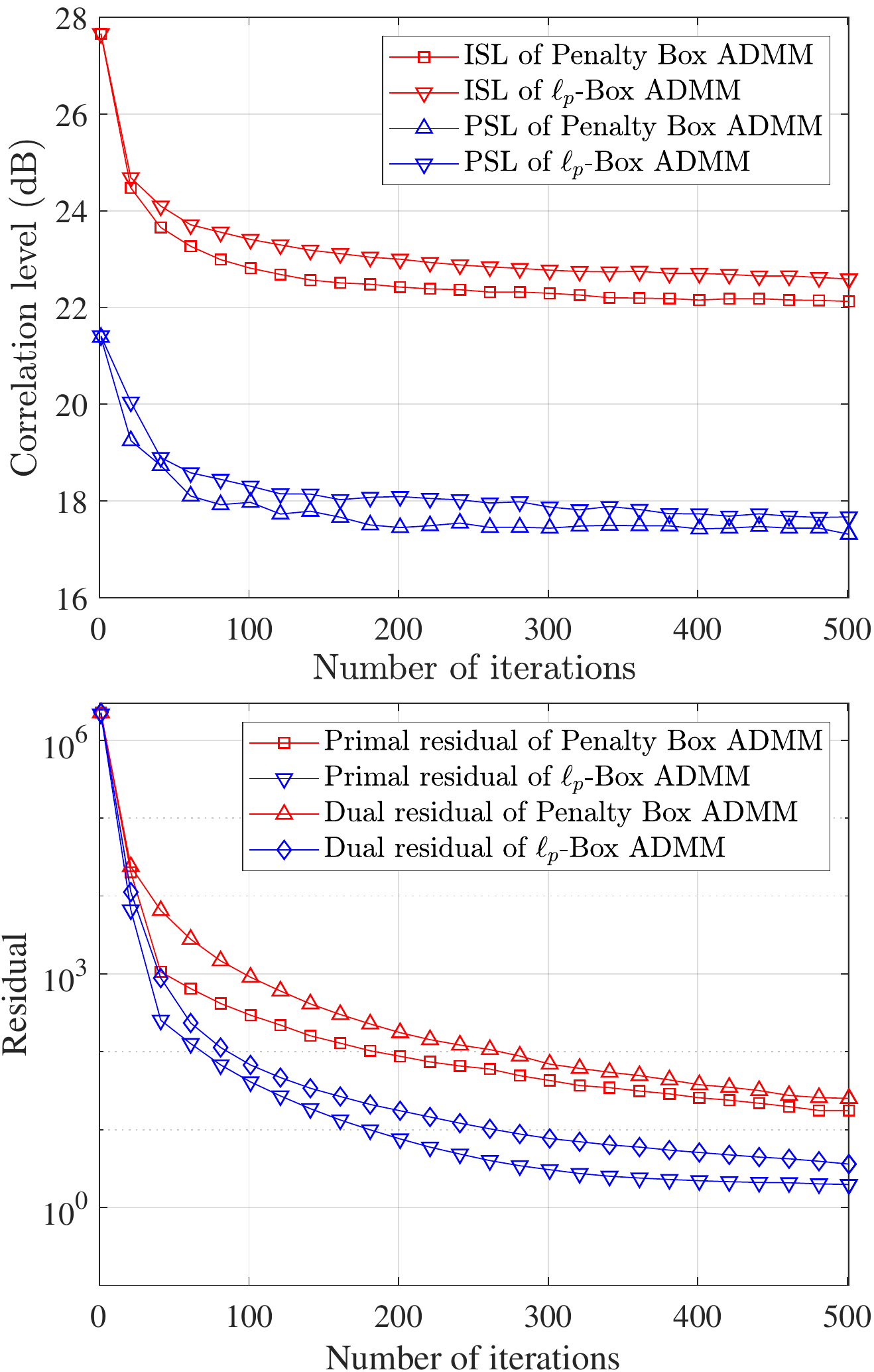}}

\caption{Comparisons of aperiodic convergence performance with $N=2048,M=1,\mathds{L}=[1,N-1]$.}
\label{conv 2048 ap}
\end{figure}

\begin{figure}[t]
\centering

\centerline{\includegraphics[scale=0.6]{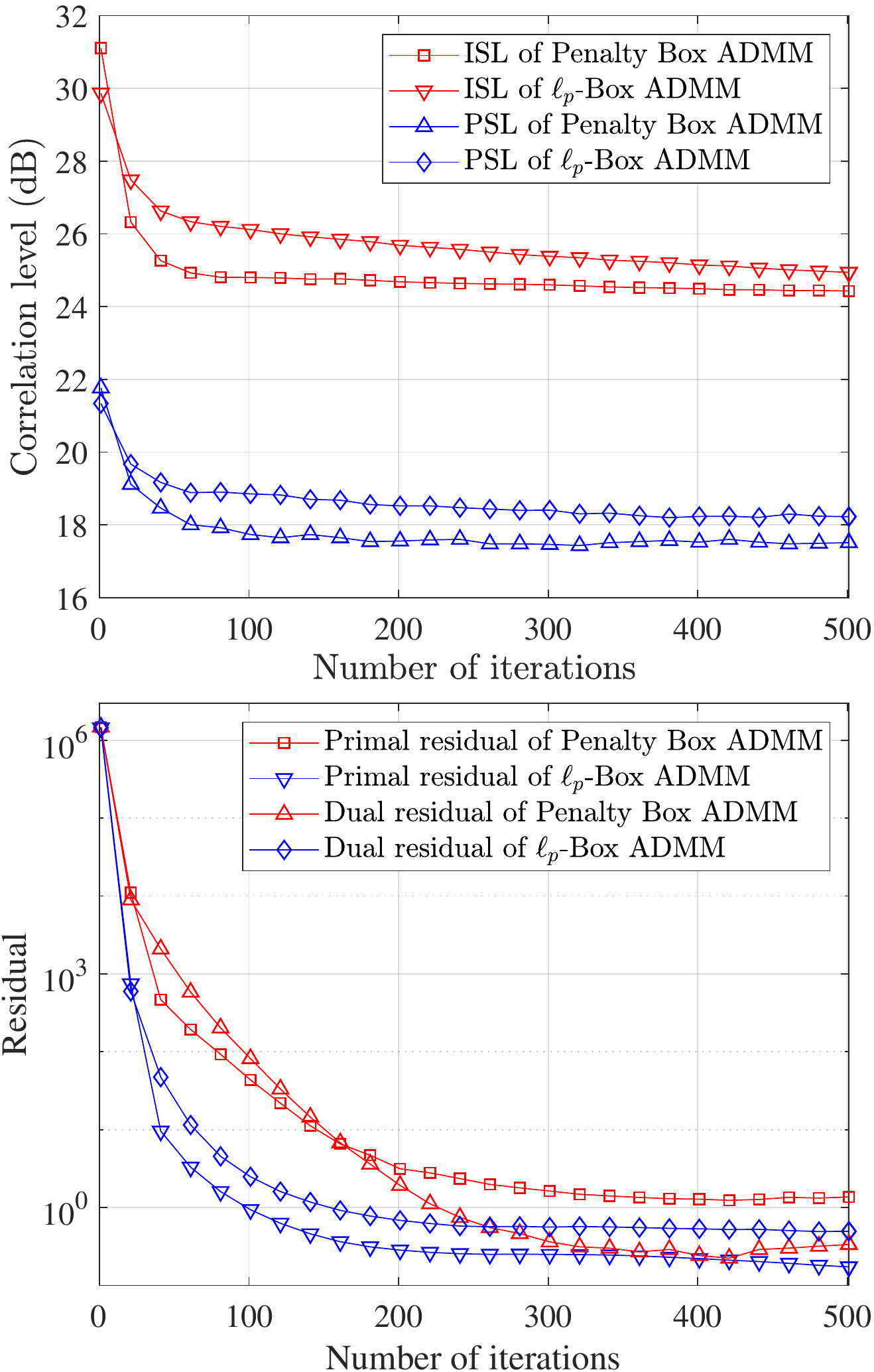}}

\caption{Comparisons of periodic convergence performance with $N=2048,M=1,\mathds{L}=[1,N-1]$.}
\label{conv 2048 p}
\end{figure}


In this section, several numerical examples are presented to illustrate the performance of the proposed ADMM algorithm.
The simulation parameters are set as follows: For the proposed ADMM algorithm, we define primal/dual residuals \cite{Boyd_11} at the $k$-th iteration as
\[
\begin{split}
\mathbf{R}^{k} = \sum\limits_{o \in \mathbb{O}}\|\mathbf{X}^{k+1} -\mathbf{Z}_o^{k+1}\|_F^2,\ \mathbf{D}^{k}= \sum\limits_{o \in \mathbb{O}}\|\mathbf{Z}_o^{k+1}-\mathbf{Z}_o^{k}\|_F^2.
\end{split}
\]
Then, the termination criterion in the simulations is set as
$
\mathbf{R}^{k}+\mathbf{D}^{k}\leq 10^{-3},
$
or the maximum iteration number $500$ is reached.
In the simulations, we set $ M=1, 2$, $L\in[6,10]$,  $ N=2^L$ (aperiodic case) or $2^L-1$ (periodic case), and $\mathds{L}= [0,N/2]$ or  $[0,N-1]$ .

In comparison, two state-of-the-art methods, BiST \cite{Kerahroodi_19} and MM-WeCorr \cite{Song_16}, are carried out here.
All approaches are initialized with the random binary sequence.
Besides, all experiments are performed in MATLAB 2019b/Windows 7 environment on a computer with 2.1GHz Intel 4100$\times$2 CPU and 64GB RAM.

\subsection{Convergence Performance}

Fig. \ref{conv 2048 ap} and \ref{conv 2048 p} show the convergence characteristics of the proposed ADMM algorithm.
All results are obtained over $50$ independent trials.
 {From Fig. \ref{conv 2048 ap}, we can see that both the ISL and PSL values of Penalty Box/$\ell_p$-Box ADMM converge faster than the Box ADMM.}
While for the residual values,  the result in  Fig. \ref{conv 2048 p}{conv 2048 p}is just the opposite.
This is mostly because during the iteration process, compared with $\ell_p$-Box ADM algorithm, the variable elements of $\mathbf{X}$ in the Box ADMM algorithm more easily tend to be non-binary solutions.
Specifically, the existence of non-binary solutions makes the residuals in each Box ADMM iteration smaller than other methods (see Fig. \ref{conv 2048 ap}).
Also it verifies that introducing extra processing methods (penalty terms or $\ell_p$-Box) to encourage binary solutions is feasible and effective.

Here, it should be noted that the proposed ADMM algorithm converge to the stationary point of the approximate problem.
The exact convergence analysis is presented in Theorem \ref{theorem converg}.





\subsection{ Aperiodic correlation Performance}

\begin{figure}[t]
\centering
\centering
\centerline{\includegraphics[scale=0.5]{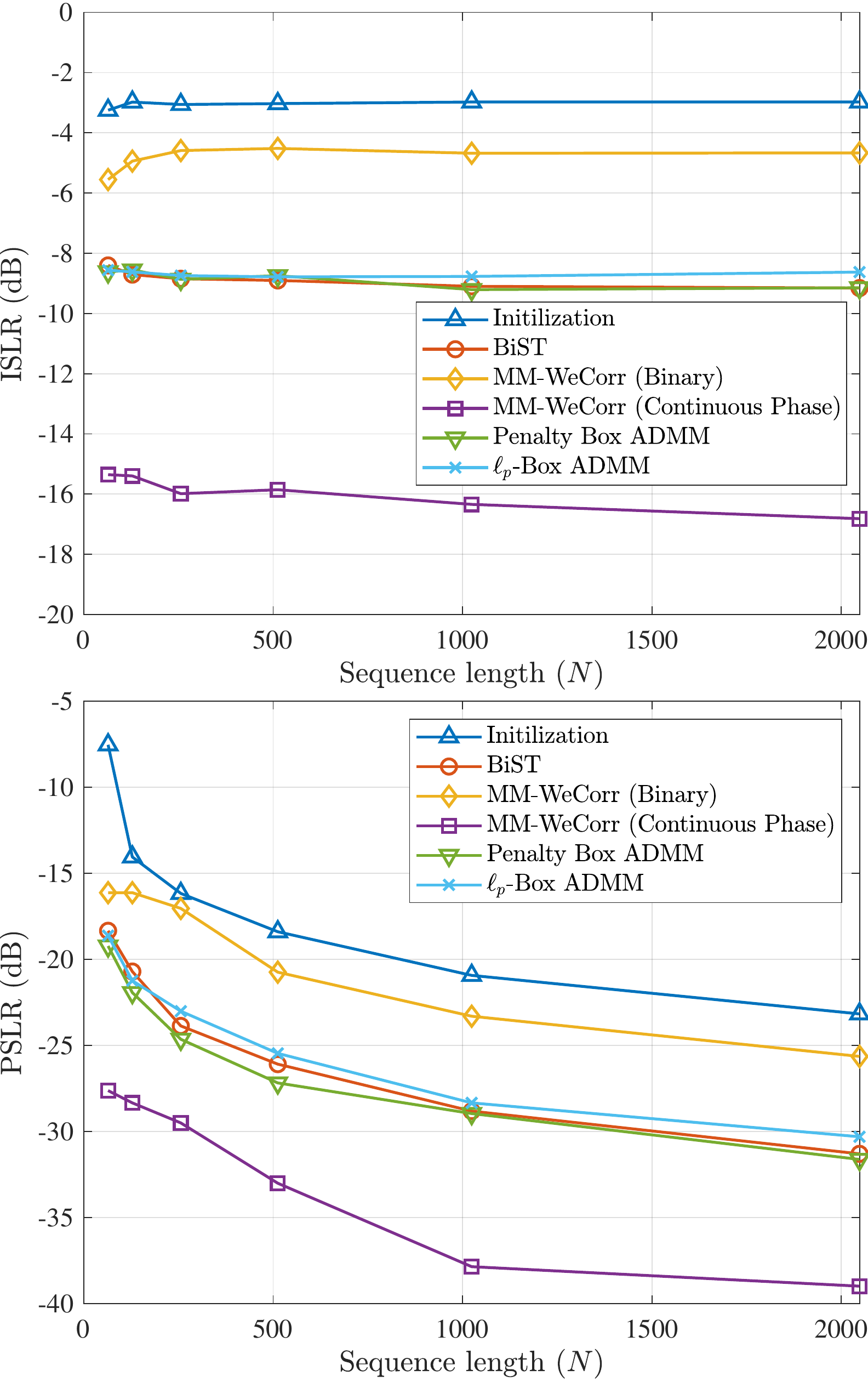}}
\caption{Comparison of the averaged aperiodic ISLR/PSLR values for different algorithms over $50$ independent trails.}
\label{corr SL}
\end{figure}
\begin{figure}[t]
\centering
\centering
\centerline{\includegraphics[scale=0.45]{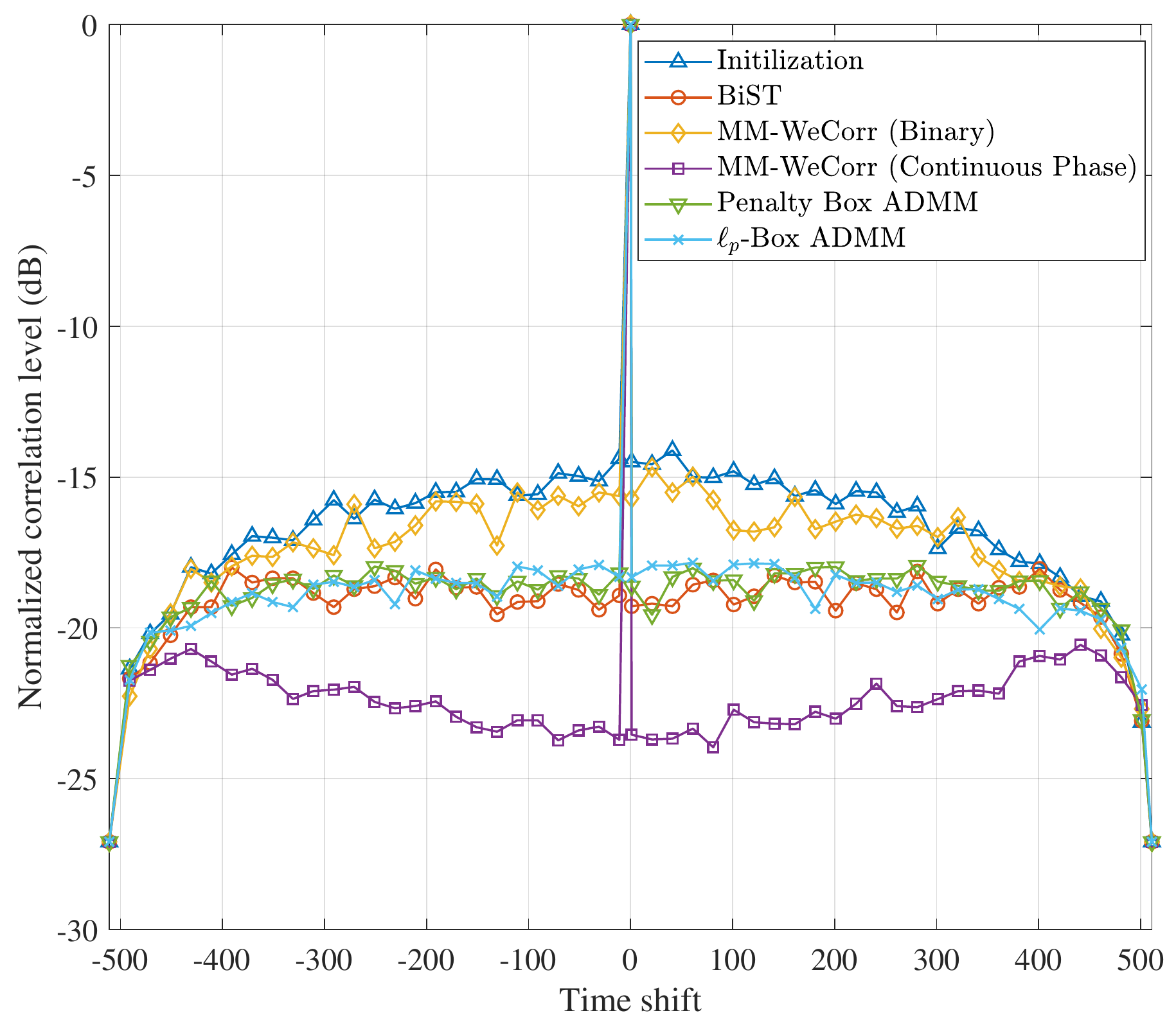}}

\caption{Aperiodic correlation levels with $N=512,M=1,\mathds{L}=[1,N-1]$.}
\label{corr 512}
\end{figure}

\begin{figure}[t]
\centering

\centering
\centerline{\includegraphics[scale=0.55]{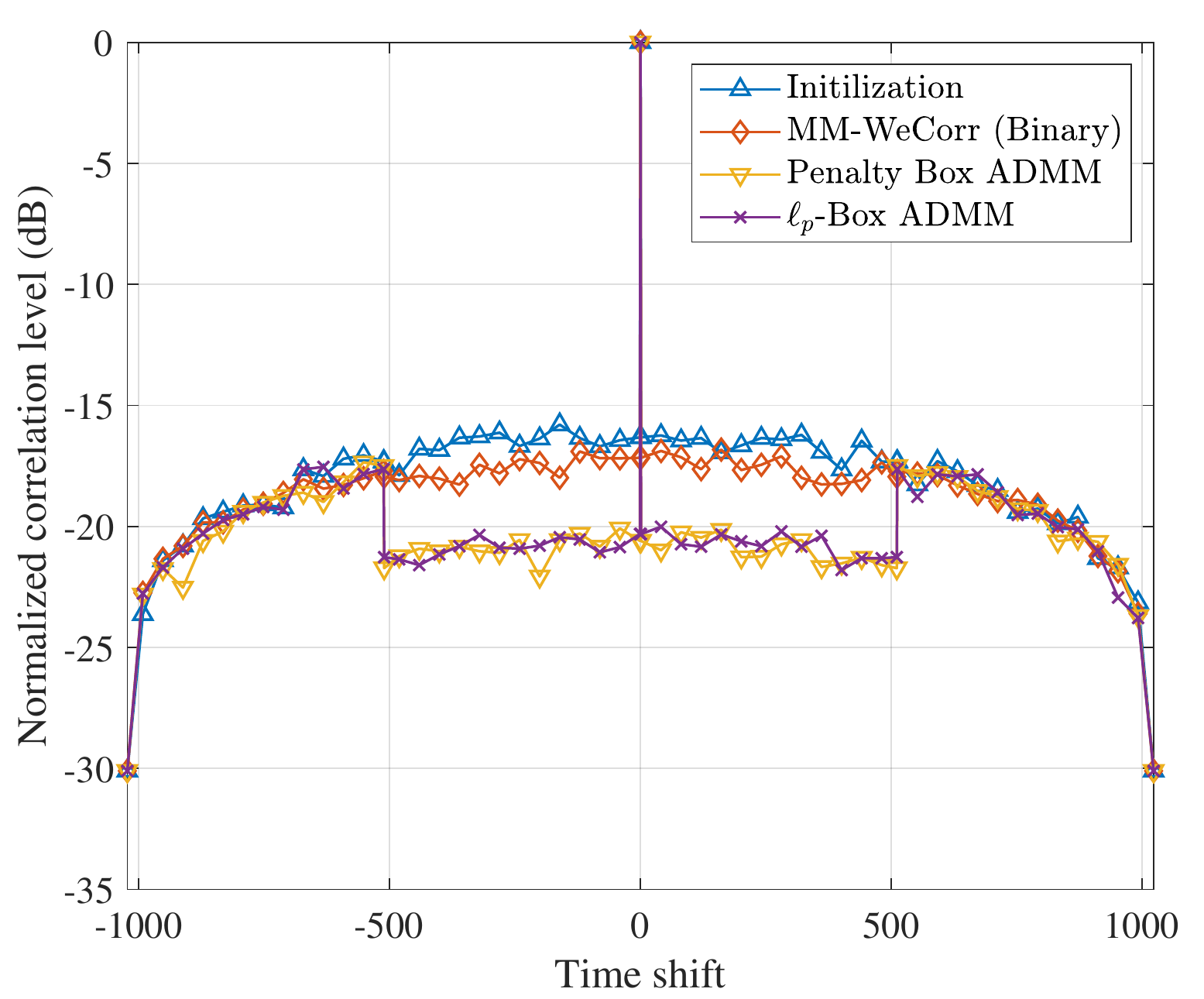}}

\caption{Aperiodic correlation levels with $N=1024,M=1,\mathds{L}=[1,N/2]$.}
\label{corr 1024}
\end{figure}

\begin{figure}[t]
\centering
\centering
\centerline{\includegraphics[scale=0.65]{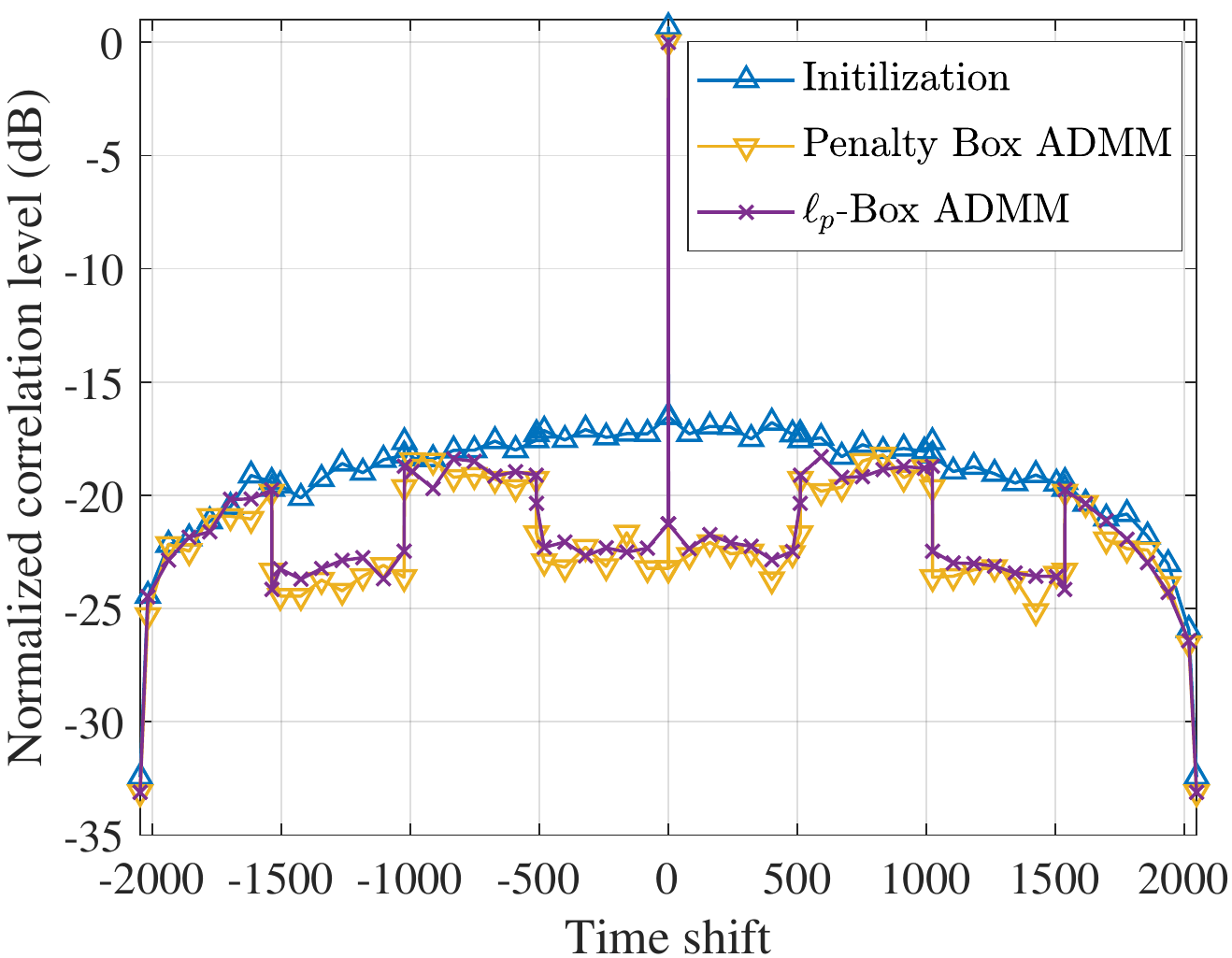}}

\caption{Aperiodic correlation levels with $N=2048, M=1, \mathds{L}=\big\{[1,N/4],[N/2,3N/4]\big\}$.}
\label{corr 2048}
\end{figure}

\begin{figure*}[t]
\centering
\centerline{\includegraphics[scale=0.72]{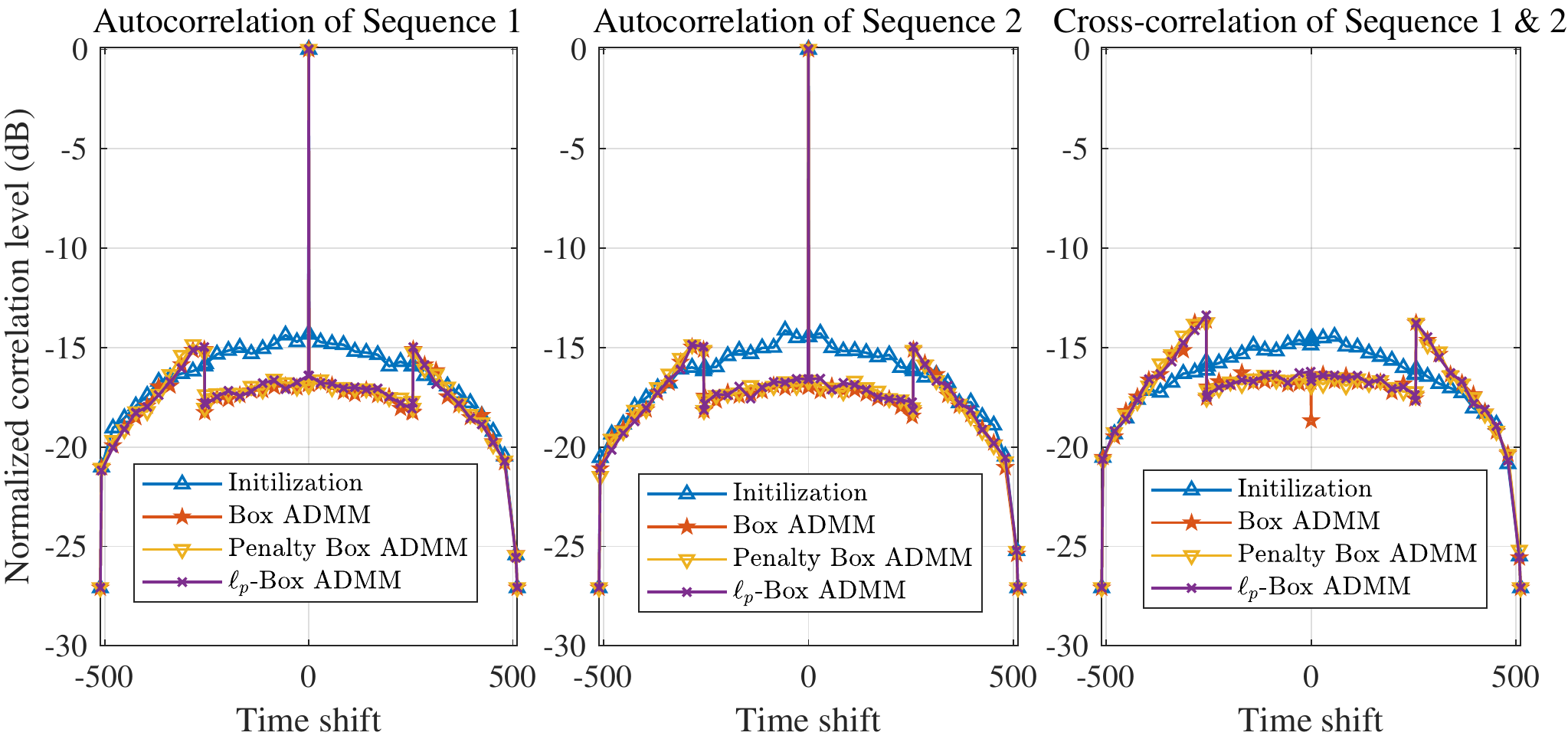}}

\caption{Aperiodic correlation levels with $N=512,M=2,\mathds{L}=[0,N/2]$.}
\label{corr 512 M2}
\end{figure*}

To evaluate the correlation properties of the binary sequences, we use the normalized ISL/PSL in $dB$
\[
\begin{split}
{\rm ISLR} = 10\lg\frac{\rm ISL}{MN^2},\ {\rm PSLR} = 10\lg\frac{{\rm PSL}^2}{MN^2}.
\end{split}
\]
Fig. \ref{corr SL}-\ref{corr 2048} compare the correlation performance between the proposed ADMM algorithm and the MM-WeCorr and BiST approach.
The sequences obtained via quantization of MM-WeCorr, named MM-WeCorr (Binary), is also considered.

From Fig. \ref{corr SL}, it can be seen that the aperiodic ISLR/PSLR of the proposed methods is similar to BiST, about $5\sim6$ dB lower than that of the initialization version.
For MM-WeCorr with continuous phase, this difference value is more than $10$ dB, which enjoys the best correlation performance for both ISLR and PSLR.
But the ISLR/PSLR of its quantified version (MM-WeCorr (Binary)) is only $2$ dB lower than the initialization version, even less.
Thus it can be seen that direct quantization of the designed continuous phase sequence has a large loss of performance, which fully explains the necessity of designing binary sequence.

Fig. \ref{corr 512} shows the example of correlation comparison with parameter $N=512,M=1,\mathds{L}=[1,N-1]$.
Among the proposed algorithms, MM-WeCorr (Binary)'s correlation performance is only better than the initialization which
As for Penalty Box/$\ell_p$-Box ADMM, their performance is similar to that of the BiST approach.

Fig. \ref{corr 1024} presented a set of experiments with shift interval $\mathds{L}= [1,N/2], N=1024,M=1$.
It should be noted that the shift interval in comparison algorithm BiST cannot be flexibly selected.
Therefore, the experiment of its algorithm is missing.
From the figure, it can be seen that compared with the random initialization, the correlation level of the proposed algorithm with $\mathds{L}= [1,N/2]$ is lower about $4$ dB, but for MM-WeCorr (Binary), only $1$ dB.
The correlation comparison with parameter $N=2048,M=1,\mathds{L}=\big\{[1,N/4],[N/2,3N/4]\big\}$ is given in Fig. \ref{corr 2048}.
Two shift intervals in $\mathds{L}$ lead MM-WeCorr (Binary) is difficult to implemented, therefore the results is missing.
The normalized correlation level in $\mathds{L}$ of proposed algorithm is lower about $4\sim5$ dB than that of random initilization.
Fig. \ref{corr 512 M2} shows the autocorrelation and cross-correlation level with $\mathds{L}= [0,N/2]$ of two sequences respectively.
From the figures, we can see that the correlation level of the sequence generated by the proposed ADMM algorithm is lower on average $2$ dB than that of the initialization sequence.
However, the same results in Fig. \ref{corr 1024} is $4\sim5$ dB with $M=1$.
The inconsistency between the results of Fig. \ref{corr 1024} and Fig. \ref{corr 512 M2} is probably caused by the lack of design degrees of freedom with the increase of $M$.

All the results in Fig. \ref{corr 1024} -- \ref{corr 512 M2} illustrate the flexibility selectable correlation interval of the model proposed in this paper.

\subsection{ Periodic correlation Performance}

\begin{figure}[t]
\centering

\centering
\centerline{\includegraphics[scale=0.6]{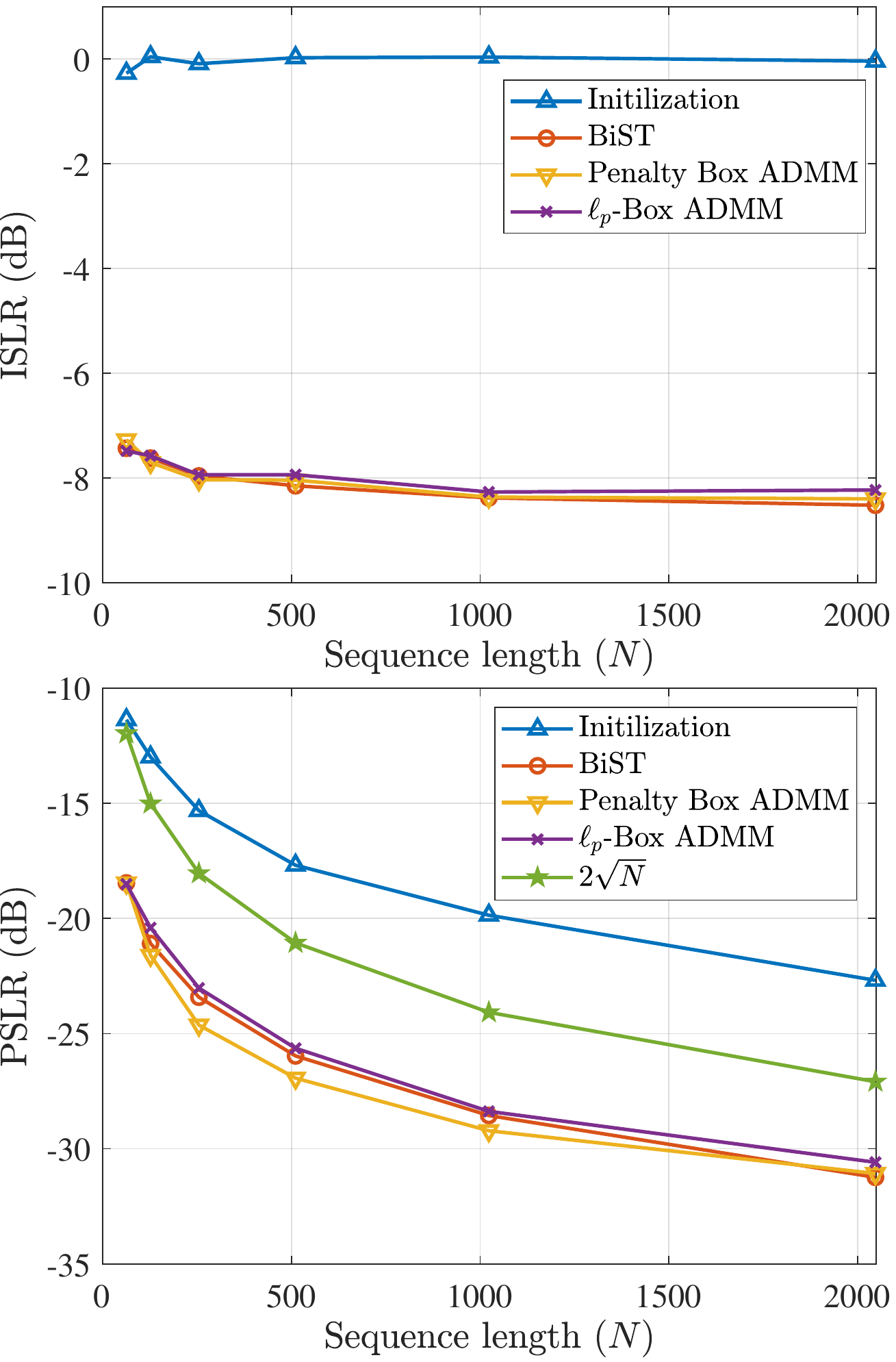}}

\caption{Comparison of the averaged periodic ISLR/PSLR values for different algorithms over $50$ independent trails.}
\label{corr 256p}
\end{figure}

\begin{figure}[t]
\centering

\centering
\centerline{\includegraphics[scale=0.5]{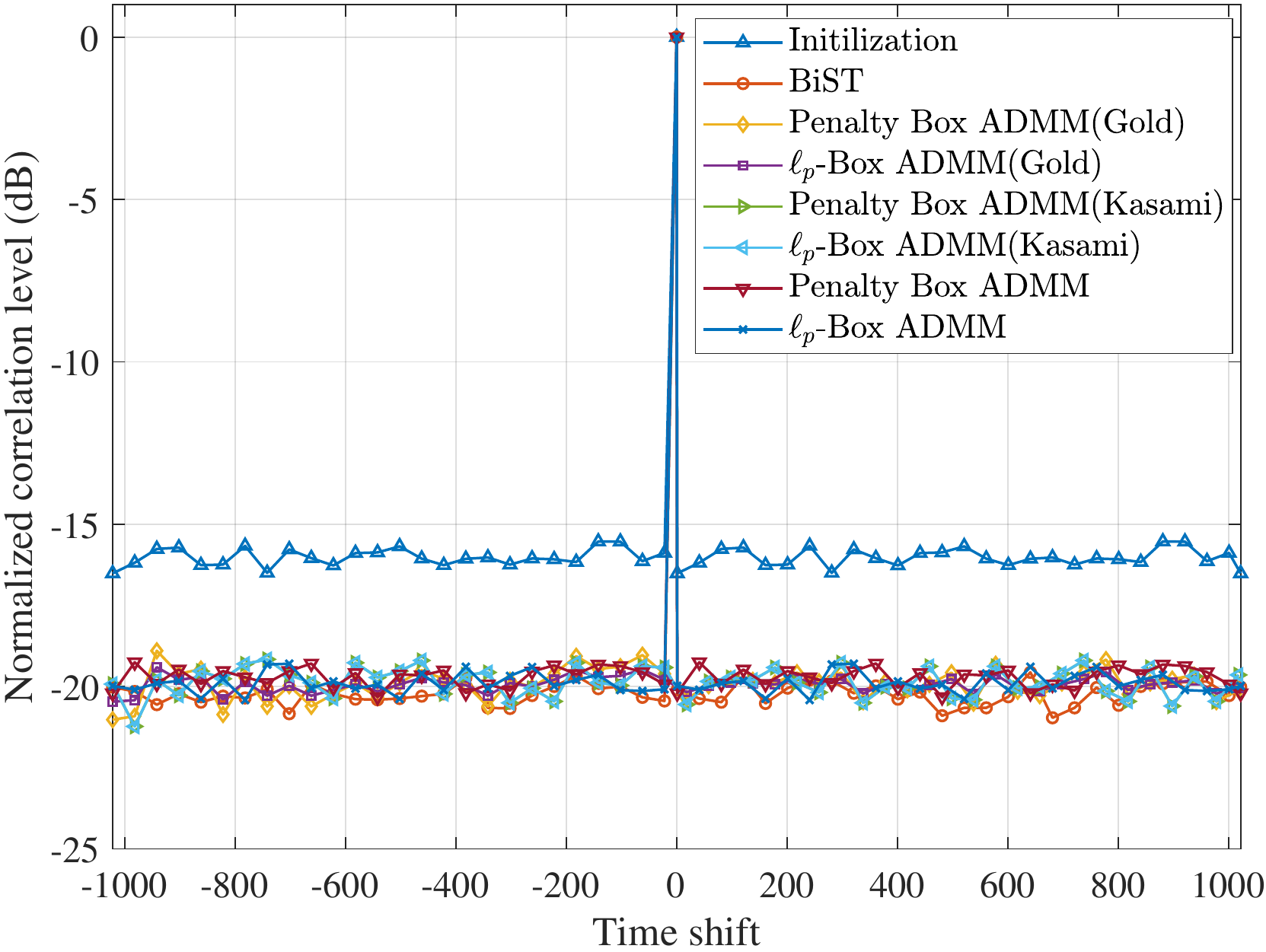}}

\caption{Periodic correlation levels with $N=1023,M=1,\mathds{L}=[1,N-1]$.}
\label{corr 1023p}
\end{figure}

Fig. \ref{corr 1023p} shows the comparison of the averaged periodic ISLR/PSLR values for different algorithms.
Reference \cite{Han_09} presented that $2\sqrt{N}$ is the PSL of the best known set of the structured sequences (i.e., Gold, Kasami, $m-$sequences, etc.).
In this paper, we consider $2\sqrt{N}$ as the periodic PSL comparison benchmark.
From the figure, it can be seen that the periodic ISLR/PSLR of the proposed $\ell_p$-Box ADMM is similar to BiST, about $7\sim8$ dB lower than that of the initialization version and $3\sim4$ dB lower than $2\sqrt{N}$.
If the weight of penalty term in Penalty Box ADMM is reasonably chosen, the periodic PSL correlation performance of the generated sequence  is a litter bit lower than that of  $\ell_p$-Box ADMM.
Since there is no penalty weight, $\ell_p$-Box ADMM is easier to implement.


\subsection{Computational Complexity}

\begin{table}[t]
\caption{{computational complexity}}
\label{complexity}
\centering
\begin{tabular}{|c|c|c|}
\hline
\hline
  {BiST} &  {MM-WeCorr} &  { proposed ADMM algorithm}   \\
\hline
 { $\mathcal{O}(MN^3)$} & {$\mathcal{O}(M^2N^2)$}  &{$\mathcal{O}(M^2N|\mathds{L}|)$ }   \\
\hline
\hline
\end{tabular}
\\
\vspace{0.1cm}
\end{table}

The computational complexity in each iteration of the proposed ADMM algorithm and two state-of-the-art algorithms is listed in Table \ref{complexity}.
Since $|\mathds{L}|\leq N$ and $M<N$, we can conclude that the computational complexity of the proposed algorithm is cheaper than BiST and MM-WeCorr.
Besides, MM-WeCorr can not be used directly to design binary sequence.
Moreover, notice that, unlike BiST and MM-WeCorr, the proposed algorithm can be performed in parallel, which means that they are more suitable for large-scale applications from a practical viewpoint of implementation.


\section{Conclusion}\label{Conclusion}

In this paper, we formulated the binary sequences with low correlation properties design problem as a nonconvex combination optimization model.
  {Then, an efficient algorithm, named by Box ADMM, was proposed to solve the formulated problem.
To encourage binary solutions,   ADMM algorithm, $\ell_p$-Box ADMM algorithm is proposed based on Box ADMM algorithm.}
We proved that, if proper parameters are chosen, the proposed ADMM algorithm converge to some stationary point of
the approximate problem.
Moreover, we also provided the computational complexity of the proposed approaches.
Numerical experiments showed that, compared to the state-of-the-art methods, the proposed algorithm can choose the optimized correlation interval flexibly and obtain good correlation performance.
Besides, the parallel implementation structure makes the proposed algorithm be more suitable for large-scale applications.

\appendices
\section{Proof of Lemma 1}\label{proof lemma lip}
The idea of proving lemma \ref{Lips cont} is based on the definition of Lipschitz continuity.
\subsection{For $\nabla f^2_{ii,0}({\mathbf{X}})$}
We have the derivations in \eqref{proof lipschitz}.
\begin{equation}\label{proof lipschitz}
\begin{split}
 \frac{\|\nabla f_{ii,0}^2({\mathbf{\tilde{X}}})-\nabla f_{ii,0}^2(\mathbf{\hat{X}})\|^2_F}{\|{\mathbf{\tilde{X}}}-\mathbf{\hat{X}}\|^2_F}
 \leq \max_{i,n}\left\{\left|\frac{\frac{\partial f_{ii,0}^2(\mathbf{\tilde{X}})}{\partial x_{i,n}}-\frac{\partial f_{ii,0}^2(\mathbf{\hat{X}})}{\partial \hat{x}_{i,n}}}{\tilde{x}_{i,n}-\hat{x}_{i,n}}\right|^2\right\}.
\end{split}
\end{equation}
Since $ f^2_{ii,0}(\mathbf{X})$ is continuous and differentiable (see \eqref{def fl}), according to the Lagrangian mean value theorem, there exists some point $\bar{x}_{i,n}\in (\tilde{x}_{i,n},\hat x_{i,n})$ which satisfies
\begin{equation}\label{mean value theorem}
\frac{\frac{\partial f_{ii,0}^2(\mathbf{\tilde{x}})}{\partial \tilde{x}_{i,n}}-\frac{\partial f_{ii,0}^2(\mathbf{\hat{x}})}{\partial \hat{x}_{i,n}}}{\tilde{x}_{i,n}-\hat{x}_{i,n}}= \frac{\partial^2 f_{ii,0}^2(\mathbf{x})}{\partial \bar{x}_{i,n}^2}.
\end{equation}
Plugging \eqref{mean value theorem} into \eqref{proof lipschitz}, we get
\begin{equation}\label{lipschitz phi}
\begin{split}
\!\!\!\!\!\!\frac{\|\nabla f_{ii,0}^2(\mathbf{\tilde{X}})\!\!-\!\!\nabla f_{ii,0}^2(\mathbf{\hat{X}})\|_F}{\|{\mathbf{\tilde{X}}}-\mathbf{\hat{X}}\|_F}\!\leq \max_{i,n}\!\left\{\!\left|\frac{\partial^2 f_{ii,0}(\mathbf{X})}{\partial \bar{x}_{i,n}^2}\right|\!\right\}\!.
\end{split}
\end{equation}
According to the definition of Lipschitz continuity, we prove Lemma \ref{Lips cont} through bounding the right-hand side of \eqref{lipschitz phi}.
Based on \eqref{partial cl}, we have the derivations in \eqref{max second derivative}.
\begin{equation}\label{max second derivative}
\begin{split}
\hspace{-0.3cm}\left|\!\frac{\partial^2 f_{ii,0}^2(\mathbf{X})}{\partial \bar{x}_{i,n}^2}\!\right|
\!\!\leq\!2 \!\left|\!\text{Tr}\left(\!\frac{ \partial f_{ii,0}}{\partial \bar{x}_{i,n}}\frac{ \partial f_{ii,0}}{\partial \bar{x}_{i,n}} \!\right)\!\right|\!\! +\!\! 2\left|\!\text{Tr}\left(\!\frac{ \partial^2 f_{ii,0}}{\partial \bar{x}_{i,n}^2}f_{ii,0} \!\right)\!\right|.
\end{split}
\end{equation}
Since $\tilde{x}_{i}\in[-c, c]$ and $\hat{x}_{i}\in[-1, 1]$, we have $\bar{x}_{i,n}\in(-\hat c,\hat c )$, where $\hat c  =\max\{c,1\}$.
From \eqref{XX}, we can see that the maximum modulus of elements in $\frac{\partial \mathbf{X}^T\mathbf{X}}{\partial{\bar x_{i,n}}}$ is ${\hat c}$ and so do it for $\frac{ \partial (\mathbf{s}_i^T\mathbf{X}^T\mathbf{X}\mathbf{s}_i)}{\partial x_{i,n}}$.
Then, the first term on the right-hand side of \eqref{max second derivative} should satisfy
\begin{equation}\label{bound first}
\begin{split}
2 \left|\text{Tr}\left(\frac{ \partial f_{ii,0}}{\partial \bar{x}_{i,n}}\frac{ \partial f_{ii,0}}{\partial \bar{x}_{i,n}} \right)\right| \leq 2{\hat c}^2.
\end{split}
\end{equation}
Since $\mathbf{X}\in\mathcal{X}_B$, the maximum modulus of elements in $\mathbf{X}^T\mathbf{X}$ is $Nc_1^2$ and \eqref{XX} indicates that  $\frac{ \partial^2 f_{ii,0}}{\partial \bar{x}_{i,n}^2}$ only has $1$ nonzero elements. Therefore, we can obtain the following inequality for the second term in \eqref{max second derivative}
\begin{equation}\label{bound second}
2\left|\text{Tr}\left(\frac{ \partial^2 f_{ii,0}}{\partial \bar{x}_{i,n}^2}f_{ii,0} \right)\right| \leq 2N{\hat c}^2.
\end{equation}
Plugging \eqref{bound first} and \eqref{bound second} into \eqref{max second derivative}, we get
\begin{equation}\label{}
\left|\frac{\partial^2 f_{ii,0}^2(\mathbf{X})}{\partial \bar{x}_{i,n}^2}\right|< 2(N+1){\hat c} ^2.
\end{equation}
Through combining the boundness of  $\left|\frac{\partial^2 f_{ii,0}^2(\mathbf{X})}{\partial \bar{x}_{i,n}^2}\right|$ with \eqref{lipschitz phi}, we can conclude that $\nabla f_{ii,0}(\mathbf{X})$ is Lipschitz continuous with the constant $L_{ii,0}\geq2(N+1){\hat c}^2$.
\subsection{For $\nabla f^2_o({\mathbf{X}})$}
Through similar above derivations, we have
\begin{equation}\label{bound phi n}
\left|\frac{\partial^2 f_o^2(\mathbf{X})}{\partial \bar{x}_{i,n}^2}\right|< 2(N+1){\hat c}^2,
\end{equation}
which results in gradients $\nabla f_o(\mathbf{X}), l\in\mathds{{L}}$ being Lipschitz continuous with constant $L_o \geq 2(N+1){\hat c}^2$.  $\hfill\blacksquare$
\section{Proof of Lemma \ref{successive difference}--\ref{lemma lower bound}}\label{lemma converg}
\subsection{Proof of Lemma \ref{successive difference}}
 To facilitate the subsequent derivations, we define the following quantities
 \begin{equation*}
\begin{split}
  \Delta _{\bar{\mathbf{X}}}^k & \!\!=\! \mathcal{L}\!\left( {{\bar{\mathbf{X}}^k},{{\left\{ {\bar{\mathbf{Z}}_o^k,{\mathbf{\Lambda }}_o^k}\! \right\}}_{o \in \mathbb{O}}}} \right)\!-\! \mathcal{L}\!\left( {{\bar{\mathbf{X}}^{k + 1}},{{\left\{ {\bar{\mathbf{Z}}_o^k,{\mathbf{\Lambda }}_o^k} \right\}}_{o \in \mathbb{O}}}} \right), \\
  \Delta _{{\bar{\mathbf{Z}}_o}}^k &\!\!= \!\mathcal{L}_{o}\!\left( \bar{\mathbf{X}}^{k + 1},\bar{\mathbf{Z}}_o^k,\mathbf{\Lambda}_o^k \right) \!-\!\mathcal{L}_{o}\!\left( {\bar{\mathbf{X}}^{k + 1}}, {\bar{\mathbf{Z}}_o^{k + 1},{\mathbf{\Lambda }}_o^k} \right), \\
  \Delta _{{{\mathbf{\Lambda }}_o}}^k &\!\!=\! \mathcal{L}_{o}\!\!\left( {{\bar{\mathbf{X}}^{k + 1}},\bar{\mathbf{Z}}_o^{k + 1},{\mathbf{\Lambda }}_o^k} \right)\!-\! \mathcal{L}_{o}\!\!\left( {{\bar{\mathbf{X}}^{k + 1}},\bar{\mathbf{Z}}_o^{k + 1},{\mathbf{\Lambda }}_o^{k + 1}} \right). \\
\end{split}
\end{equation*}
Then, the successive difference in Lemma \ref{successive difference} can be rewritten as
\begin{equation}\label{diff fun-L}
\begin{split}
\hspace{-0.2cm}& \mathcal{L}_{o}\!\left(\! {{{\mathbf{X}}^k},{{\left\{ {\bar{\mathbf{Z}}_o^k,{\mathbf{\Lambda }}_o^k} \right\}}_{o \in \mathbb{O}}}} \right) \!\!-\! \mathcal{L}_{o}\!\left( {{{\mathbf{X}}^{k + 1}},{{\left\{ {\bar{\mathbf{Z}}_o^{k + 1},{\mathbf{\Lambda }}_o^{k + 1}} \right\}}_{o \in \mathbb{O}}}}\! \right)\\
\hspace{-0.2cm}& =\Delta_{\bar{\mathbf{X}}}^k + \sum\limits_{o \in \mathbb{O}} {\left( {\Delta _{{\bar{\mathbf{Z}}_o}}^k + \Delta _{{{\mathbf{\Lambda }}_o}}^k} \right)}.
\end{split}
\end{equation}

\subsubsection{For $\Delta _{\mathbf{X}}^k$}
 {According to $\mathcal{L}_{o}(\cdot)$'s strong convexity related to $\bar{\mathbf{X}}$}, we get
\begin{equation}\label{Delta x}
\begin{split}
&\Delta_{\bar{\mathbf{X}}}^k \geq \left\langle {{\nabla _{\bar{\mathbf{X}}}}\mathcal{L}_{o}\left( {{{\bar{\mathbf{X}}}^{k + 1}},{{\left\{ {\bar{\mathbf{Z}}_o^k,{\mathbf{\Lambda }}_o^k} \right\}}_{l \in \mathbb{L}}}} \right),{{\bar{\mathbf{X}}}^k} - {{\bar{\mathbf{X}}}^{k + 1}}} \right\rangle  \\
&\hspace{2.5cm}+ \left(\sum\limits_{o \in \mathbb{O} } \frac{\rho_o}{2}\right)\left\|{\bar{\mathbf{X}}^{k + 1}} - \bar{\mathbf{X}}^k \right\|_F^2.
\end{split}
\end{equation}
Since $\bar{\mathbf{X}}^{k+1} = \underset{\mathbf{X}_1\in\mathcal{X}_B,\mathbf{X}_2\in\mathcal{X}_S} {\arg \min}\ \ \mathcal{L}_{o}\left(\bar{\mathbf{X}},\{ \bar{\mathbf{Z}}_o^k,\mathbf{\Lambda }_o^k\}_{o \in \mathbb{O}}\right)$, we have
 \begin{equation}\label{x opt}
\begin{split}
\left\langle {{\nabla _{\bar{\mathbf{X}}}}\mathcal{L}_{o}\left( {{{\bar{\mathbf{X}}}^{k + 1}},{{\left\{ {{\mathbf{Z}}_o^k,{\mathbf{\Lambda }}_o^k} \right\}}_{o \in \mathbb{O}}}} \right),{{\bar{\mathbf{X}}}^k} - {{\bar{\mathbf{X}}}^{k + 1}}} \right\rangle  \geq 0.
\end{split}
\end{equation}
Plugging it into \eqref{Delta x}, we obtain
\begin{equation}\label{Delta x end}
\begin{split}
\Delta _{\bar{\mathbf{X}}}^k \geq \left(\sum\limits_{o \in \mathbb{O} } \frac{\rho_o}{2}\right)\left\|{\bar{\mathbf{X}}^{k + 1}} - \bar{\mathbf{X}}^k \right\|_F^2.
\end{split}
\end{equation}

\subsubsection{For $\Delta _{{{\mathbf{Z}}_o}}^k$}
\begin{equation}\label{Delta z}
\begin{split}
\Delta _{{\bar{\mathbf{Z}}_o}}^k &\geq {\mathcal{L}_{o}}\left( {{\bar{\mathbf{X}}^{k + 1}}, \bar{\mathbf{Z}}_o^k,{\mathbf{\Lambda }}_o^k}   \right) - {\mathcal{U}_{o}}\left( {{{\bar{\mathbf{X}}}^{k + 1}}, \bar{\mathbf{Z}}_o^{k+1},{\mathbf{\Lambda }}_o^k}   \right)\\
&= {\mathcal{L}_{o}}\left( {{{\bar{\mathbf{X}}}^{k + 1}}, \bar{\mathbf{Z}}_o^k,{\mathbf{\Lambda }}_o^k}  \right) - {\mathcal{U}_{o}}\left( {{{\bar{\mathbf{X}}}^{k + 1}}, \bar{\mathbf{Z}}_o^k,{\mathbf{\Lambda }}_o^k}   \right)\\
&\ \ + {\mathcal{U}_{o}}\left( {{{\bar{\mathbf{X}}}^{k + 1}}, \bar{\mathbf{Z}}_o^k,{\mathbf{\Lambda }}_o^k}   \right) - {\mathcal{U}_{o}}\left( {{{\bar{\mathbf{X}}}^{k + 1}}, \bar{\mathbf{Z}}_o^{k+1},{\mathbf{\Lambda }}_o^k} \right)\\
&=\Delta _{\mathcal{L}\mathcal{U}}^k + \Delta _\mathcal{U}^k
\end{split}
\end{equation}

\begin{equation}\label{diff lu}
\begin{split}
&\ \Delta _{\mathcal{L}\mathcal{U}}^k \\
&\!=\! {{w}}_o\bigg[ {f_o^2( {{\mathbf{Z}}_o^k}) \!\!- \!\!f_o^2( {{{\mathbf{X}}^{k + 1}}})\!\! +\!\! \left\langle {\nabla f_o^2( {{{\mathbf{X}}^{k + 1}}} ),{{\mathbf{X}}^{k + 1}}\!\! -\!\! {\mathbf{Z}}_o^k} \right\rangle } \\
&\hspace{4cm}\left. { - \frac{{{L_o}}}{2}\left\| {{{\mathbf{X}}^{k + 1}} - {\mathbf{Z}}_o^k} \right\|_F^2} \right]  \\
& {\overset{(a)}{\geq}} \ {{w}}_o\bigg[ \left\langle {\nabla f_o^2\left( {{\mathbf{Z}}_o^k} \right) - \nabla f_o^2\left( {{{\mathbf{X}}^{k + 1}}} \right),{{\mathbf{X}}^{k + 1}} - {\mathbf{Z}}_o^k} \right\rangle \\
 &\hspace{4.cm}\left.- {L_o}\left\| {{{\mathbf{X}}^{k + 1}} - {\mathbf{Z}}_o^k} \right\|_F^2 \right]\\
& {\overset{(b)}{\geq}}   - 2{{w _o}} {L_o}\left\| {{{\mathbf{X}}^{k + 1}} - {\mathbf{Z}}_o^k} \right\|_F^2\\
& \geq  - 4{{w _o}} {L_o}\left\| {\frac{\mathbf{A}^T(\bar{\mathbf{X}}^{k+1}-\sigma\hat{\mathbf{Z}}_o^{k+1})}{2}- {\mathbf{Z}}_o^{k + 1}} \right\|_F^2\\
 &\hspace{4cm} - 4{{w _o}} {L_o}\left\| {{\mathbf{Z}}_o^{k + 1} - {\mathbf{Z}}_o^k} \right\|_F^2\\
& \geq - \frac{{2{{w _o}} {L_o}}}{{\rho_o^2}}\left\| {{\mathbf{\Lambda }}_o^{k + 1} \!\!-\! {\mathbf{\Lambda }}_o^k} \right\|_F^2 \!\!- \!4{{w _o}} {L_o}\left\| {{\mathbf{Z}}_o^{k + 1} \!-\! {\mathbf{Z}}_o^k} \right\|_F^2,
\end{split}
\end{equation}
where (a) and (b) come from Lemma \ref{Lips cont}.
Similar to \eqref{Delta x}, according to $\mathcal{U}(\cdot)$'s strong convexity related to $\bar{\mathbf{Z}}$, we have
\begin{equation}\label{diff u}
\begin{split}
\hspace{-0.3cm}\Delta _\mathcal{U}^k \geq \frac{{{\rho _o}\! + \!{{w _o}} {L_o}}}{2}\left\| {{\mathbf{Z}}_o^{k + 1} \!\!-\! {\mathbf{Z}}_o^k} \right\|_F^2 \!\!+ \!\frac{\rho_o \sigma^2}{2}\left\| {\hat{\mathbf{Z}}_o^{k + 1} \!-\! \hat{\mathbf{Z}}_o^k} \right\|_F^2,
\end{split}
\end{equation}
Plugging \eqref{diff lu} and \eqref{diff u} into \eqref{Delta z}, we obtain
\begin{equation}\label{Delta z end}
\begin{split}
&\hspace{-0.3cm}\Delta_{{{\mathbf{Z}}_o}}^k \!\geq\!\frac{{{\rho_o} \!-\!\!7  {{w _o}} {L_o}}}{2}\left\| {{\mathbf{Z}}_o^{k + 1}\!\! -\!\! {\mathbf{Z}}_o^k} \right\|_F^2\!\!+\!\!\frac{\rho_o \sigma^2}{2}\left\| {\hat{\mathbf{Z}}_o^{k + 1} \!-\! \hat{\mathbf{Z}}_o^k} \right\|_F^2\\
&\hspace{3.5cm} +\frac{{ -2{{w_o}} {L_o}}}{{\rho_o^2}}\!\left\| {{\mathbf{\Lambda }}_o^{k + 1} \!\!-\!\! {\mathbf{\Lambda }}_o^k} \!\right\|_F^2 .
\end{split}
\end{equation}

\subsubsection{For $\Delta _{{{\mathbf{\Lambda }}_o}}^k$}
\begin{equation}\label{Delta lambda}
\begin{split}
\Delta _{{{\mathbf{\Lambda }}_o}}^k & = {\left\langle {{\mathbf{\Lambda }}_o^{k + 1} - {\mathbf{\Lambda }}_o^k,\bar{\mathbf{A}}\bar{\mathbf{Z}}_o^{k + 1} - {\bar{\mathbf{X}}^{k + 1}}} \right\rangle } \\
& =  { - \frac{1}{{{\rho _o}}}\left\| {{\mathbf{\Lambda }}_o^{k + 1} - {\mathbf{\Lambda }}_o^k} \right\|_F^2}.
\end{split}
\end{equation}
Solving $\nabla_{\bar{\mathbf{Z}}_o}\mathcal{U}_{o}(\bar{\mathbf{Z}}_o)=0$, we get
\begin{equation}\label{lambda k}
\begin{split}
\hspace{-0.3cm}\mathbf{\Lambda}_o^{k + 1}\!\!= \!(\bar{\mathbf{A}}^T)^{-1}\!\!
\left[\!
\begin{array}{c}
\!\!{w}_o \! \left(\nabla f_o^2( {\mathbf{X}}^{k + 1}\!)  \!+\! {L_o}({\mathbf{Z}}_o^{k + 1} \!\!-\!\! {\mathbf{X}}^{k + 1}  )\right)\!\!\\
\sigma^2\hat{\mathbf{Z}}_o^{k+1}\!\!
\end{array}\!
\right].
\end{split}
\end{equation}
Plugging it into
\begin{equation}\label{diff lambda}
\begin{split}
&\left\| {{\mathbf{\Lambda }}_o^{k + 1} - {\mathbf{\Lambda }}_o^k} \right\|_F^2\\
= & \left\|\frac{1}{2}\left[
\begin{array}{ccc}
\mathbf{I}_N & \frac{1}{\sigma}\mathbf{I}_N & -\frac{1}{\sigma}\mathbf{I}_N\\
\mathbf{I}_N & -\frac{1}{\sigma}\mathbf{I}_N & \frac{1}{\sigma}\mathbf{I}_N
\end{array}
\right]\right.\\
&
\hspace{-0.6cm}\left.\cdot\!\!
\left[\!
\begin{array}{c}
 \!\!\!{{w}}_o \!\left(\! {\nabla \!f_o^2\!( {{{\mathbf{X}}^{k + 1}}}\! )\!\! -\!\! \nabla\! f_o^2\!( {{{\mathbf{X}}^k}} \!)\! \!+\! \!{L_o}\!( {{{\mathbf{Z}}_o^{k + 1}}\! \!\!- \!\!{\mathbf{X}}^{k + 1}\!\!\! -\!\! {{\mathbf{Z}}_o^k} \!\!+\!\! {\mathbf{X}}^k} \!)} \right)\!\!\!\\
\sigma^2(\hat{\mathbf{Z}}_{o1}^{k+1}-\hat{\mathbf{Z}}_{o1}^{k})\\
\sigma^2(\hat{\mathbf{Z}}_{o2}^{k+1}-\hat{\mathbf{Z}}_{o2}^{k})
\end{array}
\!\!\right]\!\!\right\|_F^2\\
\leq& w_o^2L_o^2\left(8\|{\mathbf{X}}^{k + 1}-{\mathbf{X}}^k\|_F^2+ 2 \|{\mathbf{Z}}_o^{k + 1} - {\mathbf{Z}}_o^{k}\|_F^2\right)\\
&\hspace{4.5cm} + \sigma^2\|\hat{\mathbf{Z}}_o^{k + 1}-\hat{\mathbf{Z}}_o^{k }\|_F^2.
\end{split}
\end{equation}
For $\|{\mathbf{X}}^{k + 1}-{\mathbf{X}}^k\|_F^2$, we have
\begin{equation}\label{diff xk}
\begin{split}
&\hspace{-0.3cm}\|{\mathbf{X}}^{k + 1}\!\!-\!{\mathbf{X}}^k\|_F^2 \!\!= \!\!\left\|\!\frac{\mathbf{A}^T(\bar{\mathbf{X}}^{k+1}  \!\!-\!\!\sigma\hat{\mathbf{Z}}_o^{k+1})}{2} \!\!-\!\! \frac{\mathbf{A}^T(\bar{\mathbf{X}}^{k} \!\!-\!\! \sigma\hat{\mathbf{Z}}_o^{k})}{2}\!\right\|_F^2\\
&\hspace{-0.3cm}\leq \frac{1}{4}\left(\!2\| \mathbf{A}^T(\bar{\mathbf{X}}^{k+1} \!\!- \!\!\bar{\mathbf{X}}^{k})\|_F^2 \!\! +\! 2\sigma^2\|\mathbf{A}^T(\hat{\mathbf{Z}}_o^{k+1} \!\!-\!\! \hat{\mathbf{Z}}_o^{k+1})\|_F^2\!\right) \\
& \hspace{-0.3cm}\leq \| \bar{\mathbf{X}}^{k+1} \!-\! \bar{\mathbf{X}}^{k}\|_F^2 \! + \! \sigma^2\|\hat{\mathbf{Z}}_o^{k+1} \!-\! \hat{\mathbf{Z}}_o^{k+1}\|_F^2.
\end{split}
\end{equation}
Plugging \eqref{Delta x end}, \eqref{Delta z end},  \eqref{Delta lambda}, \eqref{diff lambda} and \eqref{diff xk}  into \eqref{diff fun-L}, we get
\begin{equation*}\label{diff sum lp}
\begin{split}
\hspace{-0.1cm}&  \mathcal{L}_{o}\left(\bar{\mathbf{X}}^{k},\{ \bar{\mathbf{Z}}_o^{k},\mathbf{\Lambda }_o^{k}\}_{o \in \mathbb{O}}\right)\!-\!\mathcal{L}_{o}\left(\bar{\mathbf{X}}^{k+1},\{ \bar{\mathbf{Z}}_o^{k+1},\mathbf{\Lambda }_o^{k+1}\}_{o \in \mathbb{O}}\right)\\
\hspace{-0.3cm}&\geq \!\sum\limits_{o \in \mathbb{O}} \frac{1}{{2\rho _o^2}} \left(\bar{\epsilon}_o\left\| {{\bar{\mathbf{X}}^{k + 1}} - {\bar{\mathbf{X}}^k}} \right\|_F^2 + {\epsilon}_o \left\| {{\mathbf{Z}}_o^{k + 1} - {\mathbf{Z}}_o^k} \right\|_F^2\right.\\
&\hspace{4cm}\left. + \hat{\epsilon}_o \sigma^2\left\| {\hat{\mathbf{Z}}_o^{k + 1} - \hat{\mathbf{Z}}_o^k} \right\|_F^2\right),
\end{split}
\end{equation*}
where
\begin{equation}\label{def const}
\begin{split}
  \bar{\epsilon}_o& = \rho _o^3 - 16 {w}_o^2 L_o^2\rho _o - 32 {w}_o^3 L_o^3 , \\
  {\epsilon}_o  &= \rho _o^3 - 7 {{w _o}} {L_o}\rho _o^2 -4 {w}_o^2 L_o^2\rho_o - 8{w}_o^3 L_o^3, \\
  \hat{\epsilon}_o& = \rho _o^3 - (16{w}_o^2 L_o^2 + 2)\rho _o - 32{w}_o^3 L_o^3- 4{w}_oL_o  .
\end{split}
\end{equation}
If $\bar{\epsilon}_o,{\epsilon}_o, \hat{\epsilon}_o\geq0$,  $\mathcal{L}_{o}\left(\bar{\mathbf{X}}^{k},\{ \bar{\mathbf{Z}}_o^{k},\mathbf{\Lambda }_o^{k}\}_{o \in \mathbb{O}}\right)$ {\it decreases sufficiently}.
This completes the proof. $\hfill\blacksquare$
\subsection{Proof of Lemma \ref{lemma lower bound}}
Plugging \eqref{lambda k} into $\mathcal{L}_{o}\!\left( {{\bar{\mathbf{X}}^{k + 1}},\{\bar{\mathbf{Z}}_o^{k + 1},{\mathbf{\Lambda }}_o^{k + 1}}\}_{o \in \mathbb{O}} \right)$, we get \eqref{bound L}. $\mathcal{L}_{\mathbf{Z}_o}$ and $\mathcal{L}_{\bar{\mathbf{Z}}_o} $ can be bounded by \eqref{bound z} and \eqref{bound zh}.

\begin{figure*}
\begin{equation}\label{bound L}
\begin{split}
&\mathcal{L}_{o}\!\left( {{\bar{\mathbf{X}}^{k + 1}},\{\bar{\mathbf{Z}}_o^{k + 1},{\mathbf{\Lambda }}_o^{k + 1}}\}_{o \in \mathbb{O}} \right) \\
&=  {{w}}_o{f}_o^2\left( {{\mathbf{Z}}_o^{k + 1}} \right) + \frac{\sigma^2}{2}\hat{\mathbf{Z}}_{o}^{T}\hat{\mathbf{Z}}_{o} + \left\langle \frac{1}{2}\left[\!
\begin{array}{c}
\!\!{w}_o \! \left(\nabla f_o^2( {\mathbf{X}}^{k + 1}\!)  \!+\! {L_o}({\mathbf{Z}}_o^{k + 1} \!\!-\!\! {\mathbf{X}}^{k + 1}  )\right)\!+\!\sigma(\hat{\mathbf{Z}}_{o1}^{k+1}\!-\!\hat{\mathbf{Z}}_{o2}^{k+1})\!\!\\
\!\!{w}_o \! \left(\nabla f_o^2( {\mathbf{X}}^{k + 1}\!)  \!+\! {L_o}({\mathbf{Z}}_o^{k + 1} \!\!-\!\! {\mathbf{X}}^{k + 1}  )\right)\!+\!\sigma(\hat{\mathbf{Z}}_{o2}^{k+1}\!-\!\hat{\mathbf{Z}}_{o1}^{k+1})\!\!
\end{array}\!
\right],  \bar{\mathbf{X}}^{k+1} -\bar{\mathbf{A}}\bar{\mathbf{Z}}_o^{k+1}\right\rangle  \\
&\hspace{0.5cm} +  \frac{\rho _o}{2} \left\| \bar{\mathbf{X}}^{k+1} -\bar{\mathbf{A}}\bar{\mathbf{Z}}_o^{k+1}\right\|_F^2\\
& = \mathcal{L}_{\mathbf{Z}_o} + \mathcal{L}_{\bar{\mathbf{Z}}_o} +   \frac{\rho _o}{2} \left\| \bar{\mathbf{X}}^{k+1} -\bar{\mathbf{A}}\bar{\mathbf{Z}}_o^{k+1}\right\|_F^2\\
\end{split}
\end{equation}
\end{figure*}

\begin{figure*}
\begin{equation}\label{bound z}
\begin{split}
\mathcal{L}_{\mathbf{Z}_o}&  =  {{w}}_o{f}_o^2\left( {{\mathbf{Z}}_o^{k + 1}} \right) + \left\langle \frac{1}{2}\left[\!
\begin{array}{c}
\!\!{w}_o \! \left(\nabla f_o^2( {\mathbf{X}}^{k + 1}\!)  + {L_o}({\mathbf{Z}}_o^{k + 1} -{\mathbf{X}}^{k + 1}  )\right)\\
\!\!{w}_o \! \left(\nabla f_o^2( {\mathbf{X}}^{k + 1}\!) + {L_o}({\mathbf{Z}}_o^{k + 1} - {\mathbf{X}}^{k + 1}  )\right)
\end{array}\!\!
\right],  \bar{\mathbf{X}}^{k+1} -\bar{\mathbf{A}}\bar{\mathbf{Z}}_o^{k+1}\right\rangle \\
& = {{w}}_o{f}_o^2\left( {{\mathbf{Z}}_o^{k + 1}} \right) + \left\langle \frac{1}{2}\left[\!
\begin{array}{c}
\!\!{w}_o \! \left(\nabla f_o^2( {\mathbf{X}}^{k + 1}\!)  \!+\! {L_o}({\mathbf{Z}}_o^{k + 1} \!\!-\!\! {\mathbf{X}}^{k + 1}  )\right)
\end{array}\!\!
\right],  2{\mathbf{X}}^{k+1} -2{\mathbf{Z}}_o^{k+1}\right\rangle \\
& \overset{(a)}{\geq}  w_of_o^2\left( {\mathbf{Z}}_o^{k + 1} \right) + \left\langle w_o \nabla f_o^2\left( \mathbf{Z}_o^{k + 1} \right),\mathbf{X}^{k + 1}- \mathbf{Z}_o^{k + 1} \right\rangle -2 w_oL_o \left\| \mathbf{X}^{k + 1} - \mathbf{Z}_o^{k + 1} \right\|_F^2\\
& \overset{(b)}{\geq} {{w}}_o {f}_o^2\left( {\mathbf{X}}^{k + 1}\right) - \frac{{5{{w}}_o {L_o}} }{2}\!\left\| {\mathbf{X}}^{k + 1} - {\mathbf{Z}}_o^{k + 1} \right\|_F^2\\
& \geq {{w}}_o {f}_o^2\left( {\mathbf{X}}^{k + 1} \right) - \frac{{5{{w}}_o {L_o}} }{4}\!\left\| \bar{\mathbf{X}}^{k + 1} - \bar{\mathbf{A}}\bar{\mathbf{Z}}_o^{k + 1} \right\|_F^2,
\end{split}
\end{equation}
\end{figure*}

\begin{figure*}
\begin{equation}\label{bound zh}
\begin{split}
\mathcal{L}_{\bar{\mathbf{Z}}_o} &=  \frac{\sigma^2}{2}\hat{\mathbf{Z}}_{o}^{T}\hat{\mathbf{Z}}_{o} + \left\langle \frac{1}{2}\left[\!
\begin{array}{c}
\!\!\sigma(\hat{\mathbf{Z}}_{o1}^{k+1}\!-\!\hat{\mathbf{Z}}_{o2}^{k+1})\!\!\\
\!\!\sigma(\hat{\mathbf{Z}}_{o2}^{k+1}\!-\!\hat{\mathbf{Z}}_{o1}^{k+1})\!\!\\
\end{array}\!
\right],  \bar{\mathbf{X}}^{k+1} -\bar{\mathbf{A}}\bar{\mathbf{Z}}_o^{k+1}\right\rangle \\
& = \frac{\sigma^2}{2}\hat{\mathbf{Z}}_{o}^{T}\hat{\mathbf{Z}}_{o} + \frac{1}{2}\left\langle \sigma(\hat{\mathbf{Z}}_{o1}^{k+1}\!-\!\hat{\mathbf{Z}}_{o2}^{k+1}),(\mathbf{X}_1^{k+1}-\sigma\hat{\mathbf{Z}}_{o1}^{k+1} - {\mathbf{Z}}^{k+1} ) - (\mathbf{X}_2^{k+1}-\sigma\hat{\mathbf{Z}}_{o2}^{k+1} - {\mathbf{Z}}^{k+1} )\right\rangle\\
& \geq \frac{\sigma^2}{2}\hat{\mathbf{Z}}_{o}^{T}\hat{\mathbf{Z}}_{o} - \frac{1}{4}\left\|\sigma(\hat{\mathbf{Z}}_{o1}^{k+1}\!-\!\hat{\mathbf{Z}}_{o2}^{k+1}) \right\|_F^2 - \frac{1}{4}\left\|(\mathbf{X}_1^{k+1}-\sigma\hat{\mathbf{Z}}_{o1}^{k+1} - {\mathbf{Z}}^{k+1} ) - (\mathbf{X}_2^{k+1}-\sigma\hat{\mathbf{Z}}_{o2}^{k+1} - {\mathbf{Z}}^{k+1} )\right\|_F^2\\
& \geq - \frac{1}{4}\left(2\left\|(\mathbf{X}_1^{k+1}-\sigma\hat{\mathbf{Z}}_{o1}^{k+1} - {\mathbf{Z}}^{k+1} )\right\|_F^2 + 2\left\|(\mathbf{X}_2^{k+1}-\sigma\hat{\mathbf{Z}}_{o2}^{k+1} - {\mathbf{Z}}^{k+1} )\right\|_F^2\right)\\
& =   -\frac{1}{2} \left\| \bar{\mathbf{X}}^{k+1} -\bar{\mathbf{A}}\bar{\mathbf{Z}}_o^{k+1}\right\|_F^2,
\end{split}
\end{equation}
\hrulefill
\end{figure*}
where (a) and (b) come from  the Lipschitz continuity of the gradient of $\nabla f_o^2\left( {{{\mathbf{X}}}} \right)$.
Plugging\eqref{bound z} and \eqref{bound zh} into  \eqref{bound L} we can obtain

\begin{equation}\label{bou L}
\begin{split}
&\mathcal{L}_{o}\!\!\left( {{\bar{\mathbf{X}}^{k + 1}},\{\bar{\mathbf{Z}}_o^{k + 1},{\mathbf{\Lambda }}_o^{k + 1}}\}_{o \in \mathbb{O}} \right) \\
& \geq {{w}}_o {f}_o^2\left( {{{\mathbf{X}}^{k + 1}}} \right)\!\! +\!\! \frac{{2{\rho _o} \!\!- \!5{{w}}_o {L_o}} \!- \! 2}{4}\!\left\| {{\bar{\mathbf{X}}^{k + 1}}\!\! -\!\! \bar{\mathbf{A}}\bar{\mathbf{Z}}_o^{k + 1}} \right\|_F^2,
\end{split}
\end{equation}

Since ${w}_o{f}_o^2\left( {{{\mathbf{X}}^{k + 1}}} \right)\geq0$, if $2{\rho_o} \!\!- \!5w_oL_o \!- \! 2 \geq0$, $\mathcal{L}_{o}(\cdot)$ is lower bounded. $\hfill\blacksquare$

\section{Proof of Theorem \ref{theorem converg} }\label{proof of throrem}

\subsection{For Box ADMM algorithm}

We desire the conditions in Lemmas \ref{successive difference}-\ref{lemma lower bound} hold, i.e.,
$ \tilde{\epsilon}_o, \bar{\epsilon}_o\geq0$, and $\rho_o \geq5\bar{w}_oL_o$.
Through the famous Cardano formula \cite{Cardano}, we can get that when  {$\rho_o \geq 7.8\bar{w}_oL_o$} and $\rho_o \geq5.3\bar{w}_oL_o$, $ \tilde{\epsilon}_o, \bar{\epsilon}_o\geq0$ hold.
Therefore, we conclude that when $\rho_o \geq 7.8\bar{w}_oL_o$, all conditions in Lemmas \ref{successive difference}-\ref{lemma lower bound} hold. To simplify the description, we choose  $\forall l \in \mathds{{L}}, \rho_o \geq 8\bar{w}_oL_o$.

For $k =1,2,\cdots,+\infty $, summing both sides of \eqref{succ-diff phi}, we can obtain
\begin{equation*}
\begin{split}
\hspace{-0.3cm}&  \mathcal{L}_{o}\left(\bar{\mathbf{X}}^{1},\{ \bar{\mathbf{Z}}_o^{1},\mathbf{\Lambda }_o^{1}\}_{o \in \mathbb{O}}\right)\!-\!\mathcal{L}_{o}\left(\bar{\mathbf{X}}^{k+1},\{ \bar{\mathbf{Z}}_o^{k+1},\mathbf{\Lambda }_o^{k+1}\}_{o \in \mathbb{O}}\right)\\
\hspace{-0.3cm}&\geq \!\sum\limits_{k=1}^{+\infty}\sum\limits_{o \in \mathbb{O}} \frac{1}{{2\rho _o^2}} \left(\bar{\epsilon}_o\left\| {{\bar{\mathbf{X}}^{k + 1}} - {\bar{\mathbf{X}}^k}} \right\|_F^2 + {\epsilon}_o \left\| {{\mathbf{Z}}_o^{k + 1} - {\mathbf{Z}}_o^k} \right\|_F^2\right.\\
&\hspace{4.5cm}\left. + \hat{\epsilon}_o \sigma^2\left\| \hat{\mathbf{Z}}_o^{k + 1} - \hat{\mathbf{Z}}_o^k \right\|_F^2\right).
\end{split}
\end{equation*}
According to the boundness of $\mathcal{L}_{o}\!\left( {{\bar{\mathbf{X}}^{k + 1}},\{\bar{\mathbf{Z}}_o^{k + 1},{\mathbf{\Lambda }}_o^{k + 1}}\}_{o \in \mathbb{O}} \right)$, above equation can rewritten as
\begin{equation*}
\begin{split}
\hspace{-0.3cm}&  \mathcal{L}_{o}\left(\bar{\mathbf{X}}^{1},\{ \bar{\mathbf{Z}}_o^{1},\mathbf{\Lambda }_o^{1}\}_{o \in \mathbb{O}}\right)\\
\hspace{-0.3cm}&\geq \!\sum\limits_{k=1}^{+\infty}\sum\limits_{o \in \mathbb{O}} \frac{1}{{2\rho _o^2}} \left(\bar{\epsilon}_o\left\| {{\bar{\mathbf{X}}^{k + 1}} - {\bar{\mathbf{X}}^k}} \right\|_F^2 + {\epsilon}_o \left\| {{\mathbf{Z}}_o^{k + 1} - {\mathbf{Z}}_o^k} \right\|_F^2\right.\\
&\hspace{4.5cm}\left. + \hat{\epsilon}_o \sigma^2\left\| \hat{\mathbf{Z}}_o^{k + 1} - \hat{\mathbf{Z}}_o^k \right\|_F^2\right).
\end{split}
\end{equation*}
Since $\bar{\epsilon}_o, \epsilon_o,\hat{\epsilon}_o\geq0$ and $\mathcal{L}_{o}\!\left( {{\bar{\mathbf{X}}^{k + 1}},\{\bar{\mathbf{Z}}_o^{k + 1},{\mathbf{\Lambda }}_o^{k + 1}}\}_{o \in \mathbb{O}} \right)$ is finite, we conclude that
\begin{equation}\label{cpnv x}
\begin{split}
\lim_{k\rightarrow+\infty}\|{{\bar{\mathbf{X}}^{k + 1}} - {\bar{\mathbf{X}}^k}}\|_F =0, \ \lim_{k\rightarrow+\infty}\|{\bar{\mathbf{Z}}_o^{k + 1} - \bar{\mathbf{Z}}_o^k} \|_F= 0.
\end{split}
\end{equation}

Combining \eqref{diff lambda} and \eqref{s3 lp ADMM}, we get
\begin{equation}\label{lim result}
\begin{split}
\lim_{k\rightarrow+\infty}\|{{{\mathbf{\Lambda}}_o^{k + 1}} - {{\mathbf{\Lambda}}_o^k}} \|_F=0, \ \lim_{k\rightarrow+\infty}\|{\bar{\mathbf{X}}^{k + 1} - \bar{\mathbf{A}}\bar{\mathbf{Z}}_o^{k + 1}}\|_F = 0.
\end{split}
\end{equation}
$\left(\bar{\mathbf{X}}^{*},\{ \bar{\mathbf{Z}}_o^{*},\mathbf{\Lambda }_o^{*}\}_{o \in \mathbb{O}}\right)$ denote some limit point of the proposed ADMM algorithm.
Since $\mathbf{X}_1\in \mathcal{X}_B$, $\mathbf{X}_2\in \mathcal{X}_S$, \eqref{cpnv x} indicates $\bar{\mathbf{X}}^k$ converges to some limit point as $k\rightarrow +\infty$, i.e.,
\begin{equation}\label{lim x}
\begin{split}
 \lim_{k\rightarrow+\infty} \bar{\mathbf{X}}^{k} = \bar{\mathbf{X}}^{*}.
\end{split}
\end{equation}
Combining it with \eqref{lim result}, we can see that $\{\bar{\mathbf{Z}}_o^{k}\}$ converges.
From \eqref{lambda k}, we can obtain
\begin{equation}\label{station}
\begin{split}
\hspace{-0.3cm}\bar{\mathbf{A}}^T\mathbf{\Lambda}_o^{k + 1}\!\!= \!\!
\left[\!
\begin{array}{c}
\!\!{w}_o \! \left(\nabla f_o^2( {\mathbf{X}}^{k + 1}\!)  \!+\! {L_o}({\mathbf{Z}}_o^{k + 1} \!\!-\!\! {\mathbf{X}}^{k + 1}  )\right)\!\!\\
\sigma^2\hat{\mathbf{Z}}_o^{k+1}\!\!
\end{array}\!
\right].
\end{split}
\end{equation}
Since $\lim\limits_{k\rightarrow+\infty}{\mathbf{Z}}_o^{k } - {\mathbf{X}}^{k } =0$ and $\nabla f_o^2( {\mathbf{X}}^{k}\!)$ is Lipschitz continuous, \eqref{station} can be rewritten as
\begin{equation}\label{stationary}
\begin{split}
\hspace{-0.3cm}\bar{\mathbf{A}}^T\mathbf{\Lambda}_o^{*}\!\!= \!\!
\left[
\begin{array}{c}
\!\!{w}_o \! \nabla f_o^2( {\mathbf{X}}^{*}\!) \\
\sigma^2\hat{\mathbf{Z}}_o^{*}\!\!
\end{array}
\right].
\end{split}
\end{equation}
which means $\{\mathbf{\Lambda}_o^{k}\}$ converges.
Based on \eqref{lim result}, it can be obtained
\begin{equation}\label{sta}
\begin{split}
{\bar{\mathbf{X}}^{*} = \bar{\mathbf{A}}\bar{\mathbf{Z}}_o^{*}}.
\end{split}
\end{equation}
\eqref{stationary} and \eqref{sta} indicate that $\bar{\mathbf{Z}}_o^{*}$ is a stationary point of approximate problem\eqref{equ pro s2}.
Utilizing \eqref{def cons}, we get
\begin{equation}\label{converg}
\begin{split}
\|{\bar{\mathbf{X}}^{k + 1} - {\mathbf{A}}{\mathbf{Z}}_o^{k + 1}} \|&= \|\frac{1}{\rho_o}(\mathbf{\Lambda}_o^{k + 1}-\mathbf{\Lambda}_o^{k })+ \sigma \hat{\mathbf{Z}}_o^{k + 1}\|\\
&\leq \|\frac{1}{\rho_o}(\mathbf{\Lambda}_o^{k + 1}-\mathbf{\Lambda}_o^{k })\|+ \| \sigma \hat{\mathbf{Z}}_o^{k + 1}\|.
\end{split}
\end{equation}
Combining it with \eqref{lim result}, we can obtain
\begin{equation}\label{converg}
\begin{split}
 \lim_{k\rightarrow+\infty}\|{\bar{\mathbf{X}}^{k + 1} - {\mathbf{A}}{\mathbf{Z}}_o^{k + 1}} \| =\|{\bar{\mathbf{X}}^{*} - {\mathbf{A}}{\mathbf{Z}}_o^{*}} \|\leq \|\sigma \hat{\mathbf{Z}}_o^{*}\|=\mathcal{O}(\sigma),
\end{split}
\end{equation}
which means that the stationary point $\bar{\mathbf{Z}}_o^{*}$ is close to the stationary point of the original problem \eqref{model consensus} within the range $\mathcal{O}(\sigma)$.

$\hfill\blacksquare$

\ifCLASSOPTIONcaptionsoff
  \newpage
\fi



%
%
%




\end{document}